\documentclass[reprint,superscriptaddress,amsmath,amssymb,aps,prb,floatfix]{revtex4-1}

\usepackage{graphicx}
\usepackage{dcolumn}
\usepackage{bm}
\usepackage[colorlinks=true,allcolors=blue]{hyperref}
\usepackage{array,booktabs,tabularx}
\usepackage[caption=false]{subfig}
\usepackage[USenglish]{babel}
\usepackage{braket}

\begin{document}
\selectlanguage{USenglish}
\preprint{APS/123-QED}

\title{Phonon Collapse and Second-Order Phase Transition in Thermoelectric SnSe}

\author{Unai Aseginolaza}
\affiliation{Centro de F\'isica de Materiales CFM, CSIC-UPV/EHU, Paseo Manuel de
             Lardizabal 5, 20018 Donostia, Basque Country, Spain}
\affiliation{Donostia International Physics Center
             (DIPC), Manuel Lardizabal pasealekua 4, 20018 Donostia, Basque Country, Spain}
\affiliation{Fisika Aplikatua 1 Saila, 
             University of the Basque Country (UPV/EHU), Europa Plaza 1, 20018 Donostia,
             Basque Country, Spain}
\author{Raffaelo Bianco}
\affiliation{Dipartimento di Fisica, Universit\`a di Roma La Sapienza, Piazzale Aldo Moro 5, I-00185 Roma, Italy}
\affiliation{Graphene Labs, Fondazione Instituto Italiano di Tecnologia, Italy}
\affiliation{Department of Applied Physics and Material Science, Steele Laboratory, California Institute of Technology, Pasadena,
California 91125, United States}
\author{Lorenzo Monacelli}
\affiliation{Dipartimento di Fisica, Universit\`a di Roma La Sapienza, Piazzale Aldo Moro 5, I-00185 Roma, Italy}
\author{Lorenzo Paulatto}
\affiliation{IMPMC, UMR CNRS 7590, Sorbonne
Universit\'es - UPMC Univ. Paris 06, MNHN, IRD, 4 Place Jussieu,
F-75005 Paris, France}
\author{Matteo Calandra}
\affiliation{Sorbonne Universit\'es, CNRS, Institut des Nanosciences de Paris, UMR7588, F-75252, Paris, France}
\author{Francesco Mauri}
\affiliation{Dipartimento di Fisica, Universit\`a di Roma La Sapienza, Piazzale Aldo Moro 5, I-00185 Roma, Italy} 
\affiliation{Graphene Labs, Fondazione Instituto Italiano di Tecnologia, Italy}
\author{Aitor Bergara}
\affiliation{Centro de F\'isica de Materiales CFM, CSIC-UPV/EHU, Paseo Manuel de
             Lardizabal 5, 20018 Donostia, Basque Country, Spain}
\affiliation{Donostia International Physics Center
             (DIPC), Manuel Lardizabal pasealekua 4, 20018 Donostia, Basque Country, Spain}
\affiliation{Departamento de F\'isica de la Materia Condensada,  University of the Basque Country (UPV/EHU), 48080 Bilbao, 
             Basque Country, Spain}
\author{Ion Errea}
\affiliation{Donostia International Physics Center
             (DIPC), Manuel Lardizabal pasealekua 4, 20018 Donostia, Basque Country, Spain}
\affiliation{Fisika Aplikatua 1 Saila, 
             University of the Basque Country (UPV/EHU), Europa Plaza 1, 20018 Donostia,
             Basque Country, Spain}

\date{\today}

\begin{abstract}
Since 2014 the layered semiconductor SnSe in the high-temperature $Cmcm$ phase 
is known to be the most efficient thermoelectric material.
Making use of first-principles calculations we show that its vibrational and 
thermal transport properties are determined by huge non-perturbative 
anharmonic effects. We show that the transition from the $Cmcm$ phase
to the low-symmetry $Pnma$ is a second-order phase transition driven by the
collapse of a zone border phonon, whose frequency vanishes at the transition temperature.
Our calculations show that the spectral function of the in-plane 
vibrational modes are strongly anomalous with shoulders and double-peak structures.
We calculate the lattice thermal conductivity obtaining good agreement with experiments
only when non-perturbative anharmonic scattering is included.
Our results suggest that the good thermoelectric efficiency of SnSe is strongly affected by 
the non-perturbative anharmonicity. We expect similar effects to occur in other 
thermoelectric materials.
\end{abstract}

\maketitle

Thermoelectric materials can convert waste heat into useful 
electricity\cite{goldsmid2010introduction,behnia2015fundamentals}. 
The thermoelectric efficiency of a material is measured by the dimensionless figure of 
merit $ZT=S^{2}\sigma T/\kappa$, where $S$ is the Seebeck coefficient, $\sigma$ the electrical conductivity, 
$T$ the temperature, and $\kappa=\kappa_{e}+\kappa_{l}$ the 
thermal conductivity, constituted by electronic $\kappa_{e}$ and lattice $\kappa_{l}$ contributions.
The thermoelectric efficiency can be thus enhanced by decreasing the thermal conductivity while keeping 
a high power factor $S^{2}\sigma$. 
Materials have been doped\cite{kim2013engineered,pei2011stabilizing,heremans2008enhancement} or 
nanostructured\cite{vineis2010nanostructured,minnich2009bulk} 
in order to get a high power factor combined with a low thermal conductivity, 
yielding, i.e., $ZT\simeq 2.2$ in PbTe\cite{Hsu818}. 
In the proximity to a phase transition $ZT$ may also soar, as in the case of Cu$_{2}$Se\cite{liu2013ultrahigh}.  
Recently, however, Zhao et al. reported
for SnSe\cite{zhao2014ultralow}  the 
highest thermoelectric figure of merit ever reached in a material without doping, 
material treatment or without being in the proximity to a phase transition: $ZT\simeq 2.6$ above $800$ K.

\begin{figure}
 \includegraphics[width=\linewidth]{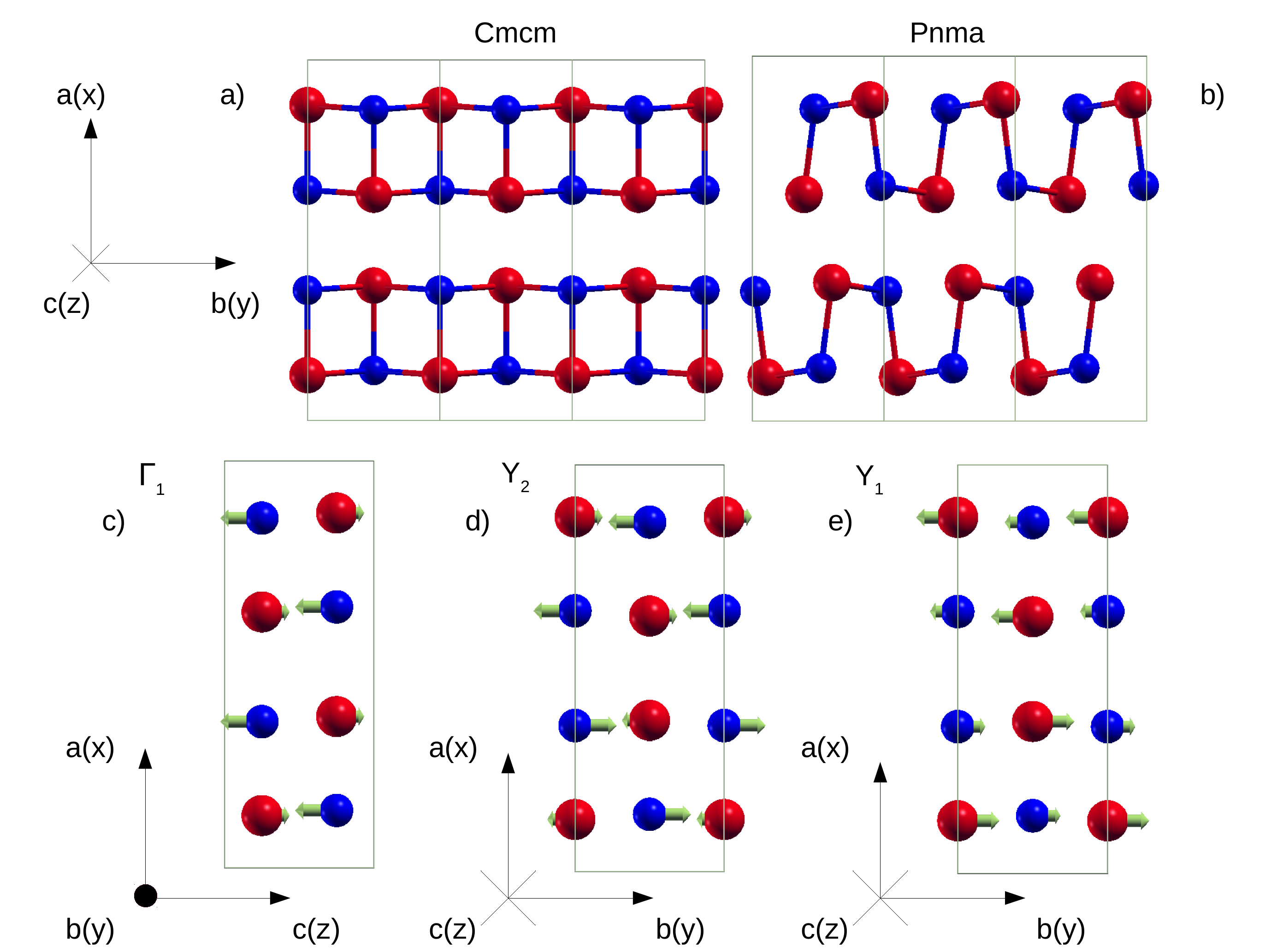}
 \caption{XY face of the a) $Cmcm$ and b) $Pnma$ structures. 
 Atomic displacements of modes c) $\Gamma_{1}$, d) $Y_{2}$ and e) $Y_{1}$. Sn  atoms 
 are red and Se blue.}
 \label{structure}
\end{figure}

SnSe is a narrow gap semiconductor that crystallizes at room temperature 
in an orthorhombic $Pnma$ phase. At  
$T\simeq800$ K\cite{zhao2014ultralow,adouby1998structure,chattopadhyay1986neutron,von1981high} 
it transforms into a more symmetric base-centered orthorhombic $Cmcm$ structure 
(see Fig. \ref{structure}).
The order of the transition is not clear: some 
works\cite{adouby1998structure,chattopadhyay1986neutron,zhao2014ultralow} claim it is a 
continuous second-order transition and 
others it has a first-order character\cite{von1981high}.
A recent work\cite{dewandre2016two} argues the transition occurs in two steps,
where increasing temperature induces first a change in the lattice parameters that induces after a
lattice instability.   
There is no inelastic scattering experiment so far for the high-temperature phase,
which should show a prominent phonon collapse at the transition temperature
if it belonged to the displacive second-order type~\cite{PhysRevLett.86.3799,PhysRevLett.107.107403,PhysRevB.95.144101}.  

The most interesting thermoelectric properties appear in the 
high-temperature $Cmcm$ phase, 
where the reduction of the electronic band gap 
increases the number of carriers providing a higher power factor, while the thermal 
conductivity remains very low\cite{zhao2014ultralow}. The value of the intrinsic 
$\kappa_{l}$ of SnSe
remains controversial, as the extremely low isotropic $0.3$ W/mK value at $800$ K
reported by Zhao et al.\cite{zhao2014ultralow} could not be reproduced in
other experiments, where a clear anisotropy is shown and
the in-plane thermal conductivity is considerably 
larger\cite{ibrahim2017reinvestigation,sassi2014assessment,chen2014thermoelectric}.
The disagreement is possibly due to large number of Sn 
vacancies in the original work\cite{zhao2014ultralow}.
The lattice thermal conductivity of the $Pnma$ phase has been calculated from first principles
solving the Boltzmann transport equation (BTE) using harmonic phonons and third order force-constants (TOFCs) 
obtained perturbatively as derivatives of the Born-Oppenheimer energy 
surface\cite{carrete2014low,skelton2016anharmonicity}. The $Cmcm$ phase
has imaginary phonon frequencies in the harmonic 
approximation\cite{dewandre2016two,skelton2016anharmonicity,yu2016enhanced}, 
as expected for the high-symmetry phase in a second order displacive 
transition\cite{iizumi1975phase,ribeiro2017anharmonic,jiang2016origin}, and it is stabilized
by anharmonicity\cite{dewandre2016two,skelton2016anharmonicity},
hindering the calculation of $\kappa_{l}$\cite{skelton2016anharmonicity}.

In this letter, by performing {\it ab initio} calculations 
fully including anharmonicity at a non-perturbative level,
we show that the phonon mode that drives the instability collapses at the 
transition temperature $T_{c}$ demonstrating that the transition is second-order. 
Anharmonic effects are so large that the spectral function
expected for some in-plane modes deviates from the simple Lorentzian-like
shape and shows broad peaks, shoulders and satellite peaks, as in 
other monochalcogenides\cite{Li2014phonon,ribeiro2017anharmonic}.
We calculate the lattice 
thermal conductivity of the $Cmcm$ phase by combining the 
anharmonic phonon spectra with perturbative and non-perturbative
TOFCs. We show here for the first time
that non-perturbative anharmonic effects are not only crucial in the
phonon spectra, but also in high-order force-constants, which
here have a huge impact on the calculated thermal conductivity:
$\kappa_l$ agrees with experiments\cite{ibrahim2017reinvestigation} only with 
non-perturbative TOFCs. Similar strong non-perturbative effects are thus expected
for other thermoelectric compounds.

The group/subgroup index of the $Cmcm$/$Pnma$ transition is 2,
making a displacive second-order transition possible\cite{toledano1987landau}. 
In this scenario, the transition temperature $T_{c}$ is defined 
as the temperature at which the second derivative of 
the free energy $F$ with respect to the order parameter $Q$
that transforms the structure continuously from the $Cmcm$ phase ($Q=0$) 
into the $Pnma$ ($Q\ne0$) vanishes. 
Symmetry~\cite{Orobengoa:ks5225,Perez-Mato:sh5107}
dictates that the amplitude of the transition is dominated 
by the distortion pattern associated to a non-degenerate mode ($Y_{1}$) 
at the zone border Y point
with irreducible representation $Y_2^+$ (see Fig. \ref{structure} 
for the distortion pattern). This means that 
$\partial^{2}F/\partial Q^{2}$ is proportional 
to the eigenvalue of the free energy Hessian matrix
associated to this irreducible representation: $\omega^2_{Y_{1}}$. 

In this work we calculate the free energy Hessian matrix
using the stochastic self-consistent harmonic 
approximation (SSCHA)\cite{errea2014anharmonic,bianco2017second,2018arXiv180406793M},
which is applied making use of \emph{ab initio} density-functional 
theory (DFT) calculations within the Perdew-Burke-Ernzerhof (PBE)\cite{perdew1996generalized} 
or Local Density Approximation (LDA)\cite{perdew1981self} parametrizations of the 
exchange-correlation functional (see Supplementary Material for the details of the  
calculations\cite{baroni2001phonons,paulatto2013anharmonic,giannozzi2009quantum,0953-8984-29-46-465901,ShengBTE_2014}).
The SSCHA is based on variational minimization of the free energy making use of a trial harmonic density 
matrix $\rho_{\mathbf{\mathcal{R}},\mathbf{\Phi}}$ 
paramatrized by centroid positions $\boldsymbol{\mathcal{R}}$ and force-constants 
$\mathbf{\Phi}$ (bold symbols represent in compact notation vectors or tensors). The centroid 
positions $\boldsymbol{\mathcal{R}}$ determine the most probable position of the atoms and
$\mathbf{\Phi}$ is related to the amplitude of their fluctuations around $\boldsymbol{\mathcal{R}}$.
The free energy Hessian can be calculated as\cite{bianco2017second}
\begin{equation}
 \frac{\partial^{2}F}{\partial\boldsymbol{\mathcal{R}}\partial\boldsymbol{\mathcal{R}}}=\mathbf{\Phi} + 
\overset{(3)}{\mathbf{\Phi}}\mathbf{\Lambda}(0)[\mathbf{1}-\overset{(4)}{\mathbf{\Phi}}\mathbf{\Lambda}(0)]^{-1}\overset{(3)}{\mathbf{\Phi}},
\label{second_derivative}
\end{equation}
where $\overset{(3)}{\mathbf{\Phi}}$ and $\overset{(4)}{\mathbf{\Phi}}$ are third- 
and fourth-order non-perturbative force-constants obtained as quantum averages calculated with 
$\rho_{\mathbf{\mathcal{R}},\mathbf{\Phi}}$:
$\overset{(n)}{\mathbf{\Phi}}=\left\langle\frac{\partial^{n}V}{\partial\mathbf{R}^n}\right\rangle_{
 \rho_{\mathbf{\mathcal{R}},\mathbf{\Phi}}}$.
The $\overset{(n)}{\mathbf{\Phi}}$ force-constants are generally different from the 
$\overset{(n)}{\mathbf{\phi}}$ perturbative ones
obtained as derivatives of the Born-Oppenheimer potential $V$ at the minimum:
$\overset{(n)}{\boldsymbol{\phi}}= \left[ \frac{\partial^{n}V}{\partial\mathbf{R}^n}\right]_{0}$. 
$\mathbf{\Lambda}(0)$ in Eq. \eqref{second_derivative} is 
a function of the $\tilde{\Omega}_{\mu}$ SSCHA frequencies and polarization vectors obtained
diagonalizing $\Phi_{ab}/\sqrt{M_aM_b}$, 
with $M_a$ the atomic mass ($a$ labels both an atom and Cartesian index). 
The $\omega_{\mu}$ frequencies obtained instead from the free energy Hessian after diagonalizing 
$\frac{\partial^{2}F}{\partial\mathcal{R}_a\partial\mathcal{R}_b}/\sqrt{M_aM_b}$, e.g. $\omega_{Y_{1}}$, 
can be interpreted as the static limit of the physical phonons\cite{bianco2017second}. 
The contribution of $\overset{(4)}{\boldsymbol{\Phi}}\boldsymbol\Lambda$ is negligible with respect to the 
identity matrix (see Supplementary Material) and thus it is neglected throughout.

\begin{figure}[th]
\includegraphics[width=\linewidth]{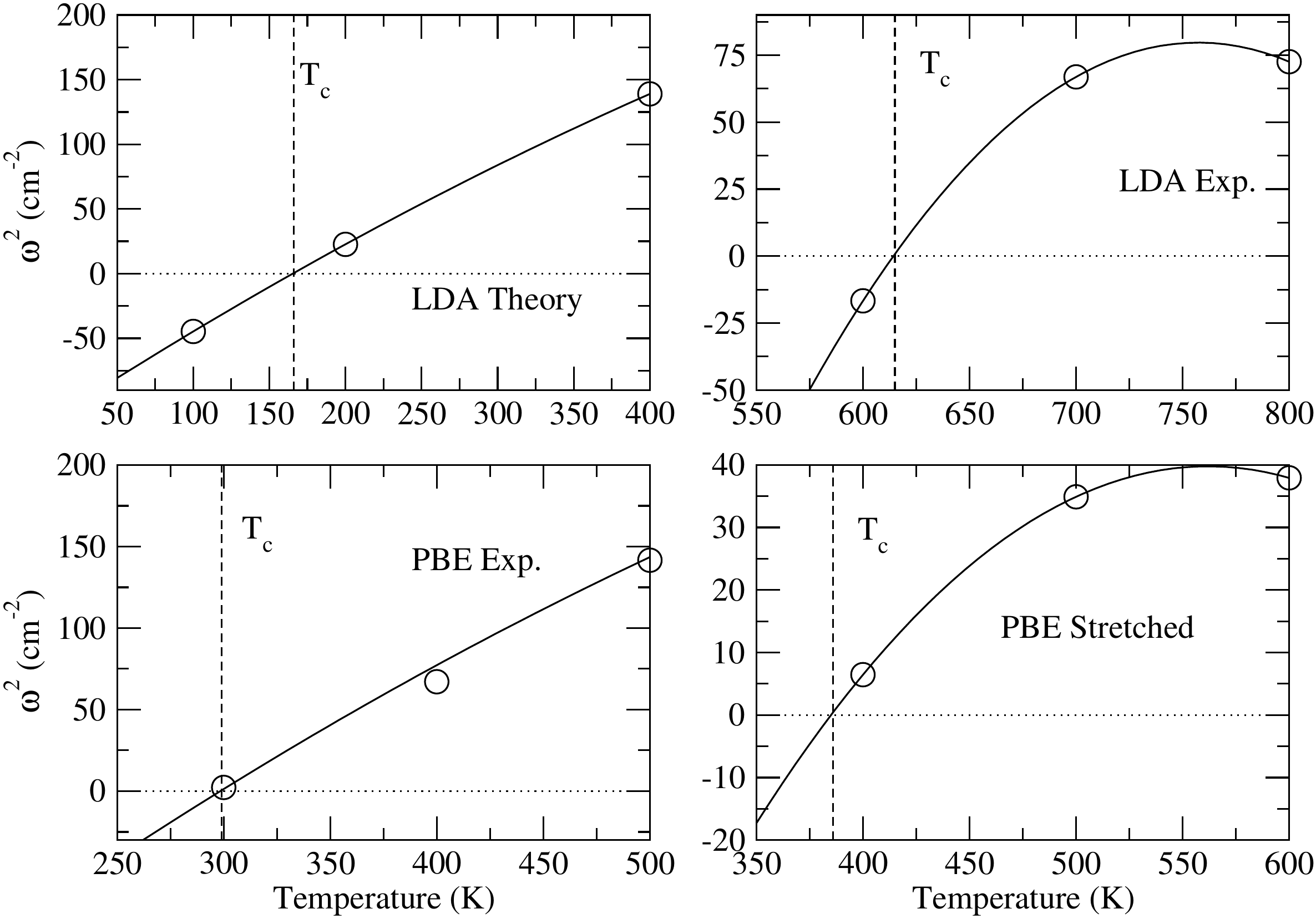}
\caption{$\omega^{2}_{Y_1}$ as a function of temperature within LDA and PBE approximations for different lattice volumes (circles). 
In the LDA we compare the results obtained with the theoretical and experimental\cite{zhao2014ultralow} lattice parameters.
In the PBE calculation we present the results for the experimental lattice parameters 
and a stretched unit cell (see Table \ref{lattice}
to check the lattice parameters in each case). The solid lines correspond to a polynomial fit.} 
\label{frequenciess}
\end{figure}

The calculated temperature dependence of $\omega^2_{Y_1}$ is shown in Fig. \ref{frequenciess}
for LDA and PBE for two different lattice volumes in each case. 
In all cases $\omega_{Y_1}^{2}$  is positive at high temperatures, but it rapidly 
decreases with lowering the temperature, vanishing at $T_c$.
This phonon collapse is consistent with a second-order phase transition between the $Pnma$ and $Cmcm$. 
We indeed check that a SSCHA calculation at $T>T_{c}$ ($T=800$ K) starting from 
the relaxed low-symmetry $Pnma$ phase yields the high-symmetry $Cmcm$ atomic positions
for the $\boldsymbol{\mathcal{R}}$ centroids. Thus, the $Pnma$ is not a local minimum
of the free energy above $T_c$, completely ruling out the first-order transition.
Our result is at odds with the
conclusions drawn in Ref. \onlinecite{dewandre2016two}.
First, because at the $T_{c}$ calculated in Ref. \onlinecite{dewandre2016two}, 
which is estimated by comparing the
free energies of the two structures, the $Y_1$ mode of the $Cmcm$
phase is stable, which implies this phase is a local minimum
at $T_{c}$, and, thus, the transition is of
first-order type\cite{dewandre2016two}. And second,
because it is argued\cite{dewandre2016two} that the instability at $Y$ 
is produced by a slight change in the in-plane lattice parameters induced by temperature 
(from $b/c<1$ to $b/c>1$), which makes the transition a two-step process.  
We do not see this sudden appearance of the instability, the $Y_1$ mode is always unstable at the harmonic
level even exchanging $b$ and $c$ (see Supplementary Material). 

\begin{table}
\begin{center}
\begin{tabular*}{0.4\textwidth}{l c c c c c c}
 \hline
 \hline
            & $a$  & $b$  & $c$  & $P_{xx}$ & $P_{yy}$ & $P_{zz}$ \\
 \hline
 LDA theory                  &  21.58  &  7.90  & 7.90  &  0.4  &  0.7  &  0.6  \\
 LDA Exp.                    &  22.13  &  8.13  & 8.13  & -1.1  & -2.2  & -2.0  \\
 PBE theory                  &  22.77  &  8.13  & 8.13  &  0.5  &  1.0  &  1.1  \\
 PBE Exp.                    &  22.13  &  8.13  & 8.13  &  1.8  &  1.2  &  1.3  \\
 PBE Stretched               &  23.48  &  8.27  & 8.27  & -0.3  & -0.7  & -0.7  \\
 \hline
 \hline
\end{tabular*}
\end{center}
\caption{Experimental\cite{zhao2014ultralow} and theoretical (DFT at static level) 
LDA and PBE lattice parameters used in this work. The stretched cell used in some 
calculations is also given. $a$, $b$, and $c$ latice parameters are given in Bohr 
length units ($a_0$) and the three components of the stress tensor in GPa units. 
The pressure is calculated including vibrational terms at an anharmonic level at 
the following temperatures for each case: $200$ K (LDA theory), $600$ K (LDA Exp.), 
$400$ K (PBE Exp.), $400$ K (PBE theory), and $400$ K (PBE stretched).} 
\label{lattice}
\end{table}

\begin{figure}[th]
\includegraphics[width=\linewidth]{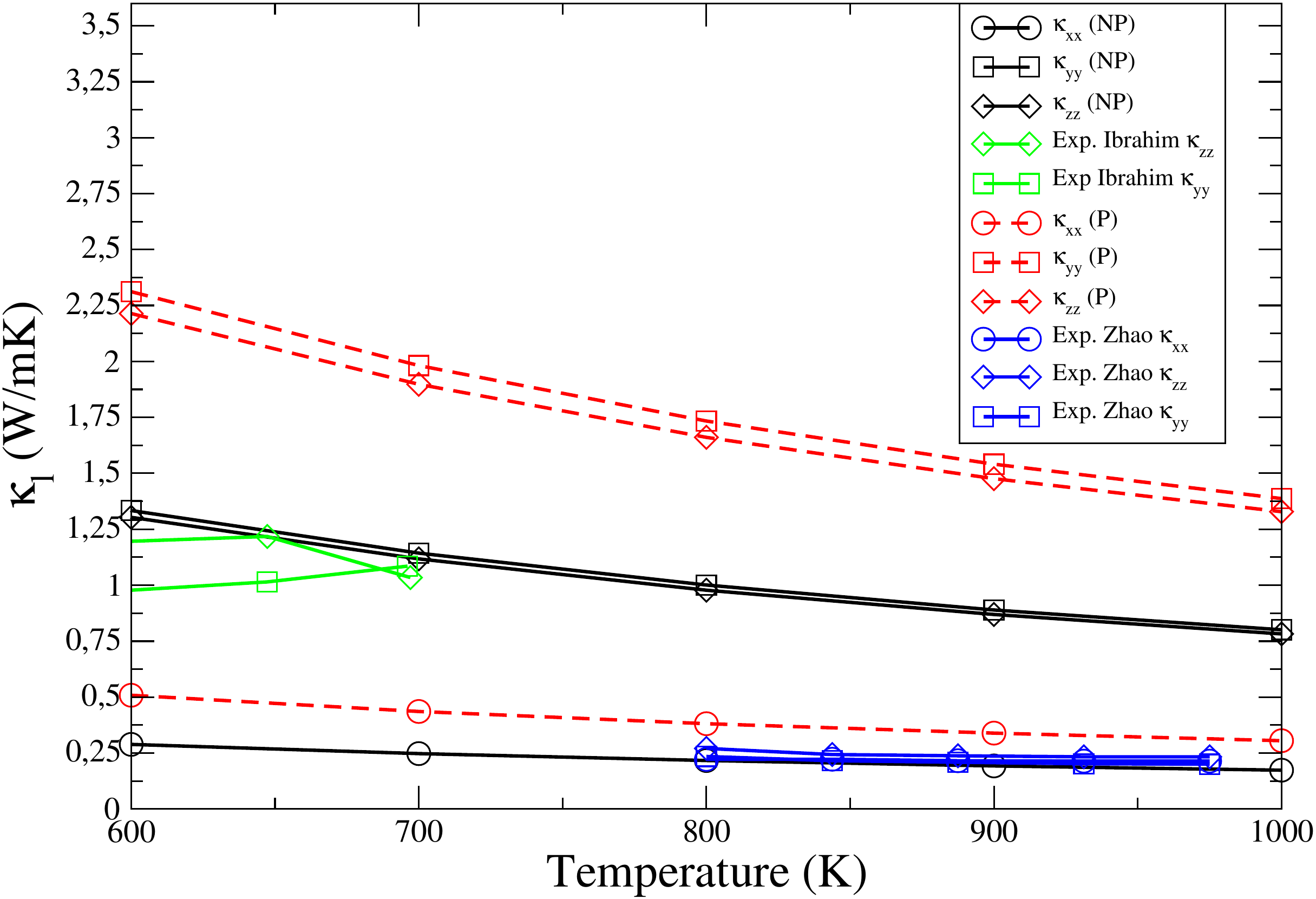}
\caption{Lattice thermal conductivity of SnSe calculated with perturbative 
$\overset{(3)}{\boldsymbol{\phi}}$ (P) and 
non-perturbative $\overset{(3)}{\boldsymbol{\Phi}}$ (NP) at $800$ K  
TOFCs using the SMA compared to the experiments 
by Ibrahim et al.\cite{ibrahim2017reinvestigation} and Zhao et al.\cite{zhao2014ultralow}. 
We use the $\tilde{\Omega}_{\mu}(\mathbf{q})$ phonon
frequencies calculated at 800 K at all temperatures. 
Calculations are performed within LDA using the experimental structure. Different volumes 
or exchange-correlation functionals give consistent results (see Supplementary Material).}
\label{experiments}
\end{figure}

The obtained transition temperature strongly depends on the exchange-correlation functional and the volume. 
Within LDA $T_{c}$ 
ranges between $168$ K with theoretical lattice parameters 
and $616$ K  with experimental lattice parameters\cite{zhao2014ultralow}
(see Table \ref{lattice} for the lattice parameters).
Within PBE $T_c$ barely changes between the experimental and theoretical
lattice parameters. We attribute this result to the fact that 
the in-plane lattice parameters $b$ and $c$ are in perfect agreement
with the experimental results within PBE, while LDA clearly underestimates 
them. 
The theoretical lattice parameters are estimated neglecting vibrational contributions
to the free energy. 
In order to estimate the role of the thermal expansion,
we calculate the stress tensor including vibrational contributions at the anharmonic
level following the method recently developed by Monacelli 
et al.\cite{2018arXiv180406793M}. The in-plane contribution
of the stress tensor calculated at the temperature closest to $T_{c}$, $P_{yy}$, shows that 
both theoretical LDA and PBE lattices should be stretched. 
In the LDA case it is clear that stretching the lattice increases
$T_c$. In the PBE case, when we take a stretched lattice to reduce $P_{yy}$, 
$T_c$ increases from $299$ K to $387$ K. In all cases the other in-plane component of the stress tensor, $P_{zz}$, is 
very similar to $P_{yy}$. 
The LDA transition temperature with the experimental lattice parameters yields the
transition temperature in closest agreement with experiments, which is estimated 
to be of $\simeq800$ K\cite{li2015orbitally,zhao2014ultralow,adouby1998structure,chattopadhyay1986neutron,von1981high}.
The underestimation of the transition temperature may be due to the approximated exchange-correlation
or the finite supercell size taken for the SSCHA. 
It is interesting to note that in the experiments by Ibrahim et al.\cite{ibrahim2017reinvestigation},
where it is stated that the samples contain much less Sn vacancies than in
Ref. \onlinecite{zhao2014ultralow}, 
the in-plane  thermal conductivity seems to increase at around $600$ K,
which may be a fingerprint of a phase transition (see Fig. \ref{experiments}). 

\begin{figure}[t]
\includegraphics[width=\linewidth]{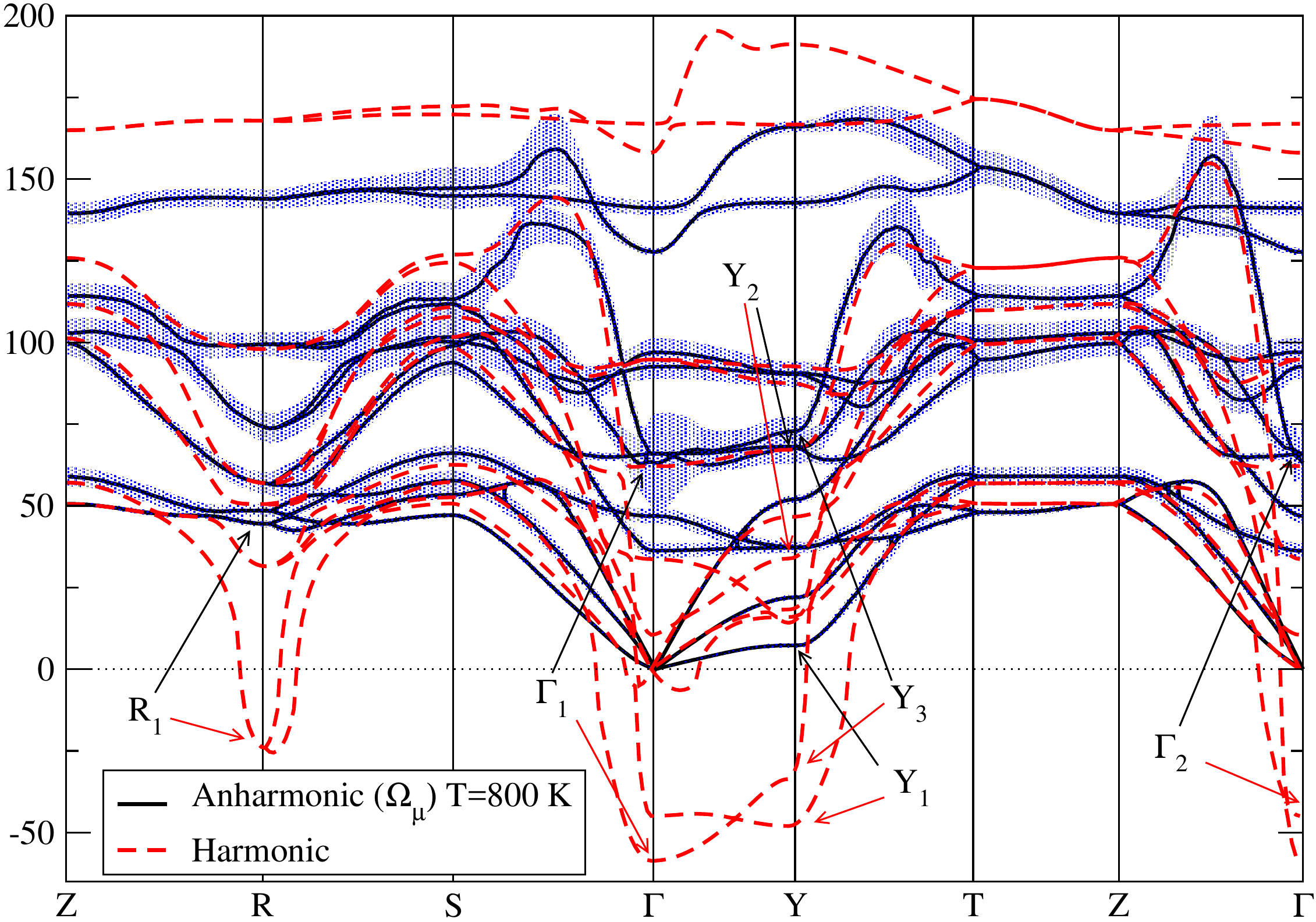}
\caption{Harmonic and anharmonic phonons in the Lorentzian approximation ($\Omega_{\mu}(\mathbf{q})$).
The length of the bars corresponds to the linewidth (full length of the line is 
the full width at half maximum). 
The calculations are done within LDA in the experimental structure using 
$\overset{(3)}{\mathbf{\Phi}}$ at $800$ K and $\tilde{\Omega}_{\mu}(\mathbf{q})$ at $800$ K.}
\label{spectrum}
\end{figure}

The phonon collapse predicted here should be experimentally measurable by inelastic neutron scattering (INS)  
experiments. INS experiments\cite{li2015orbitally} show a softening of a zone-center optical mode
of the $Pnma$ phase upon heating, which is consistent with the condensation of the
$Y_1$ mode after the transition. By making use of a dynamical ansatz\cite{bianco2017second}, 
we calculate the mode-projected phonon anharmonic self-energy $\Pi_{\mu}(\mathbf{q},\omega)$ 
(see Supplementary Material),
from which we obtain 
the phonon spectral function:
\begin{multline}
 \sigma(\mathbf{q},\omega)=\frac{1}{\pi} \times \\ \sum_{\mu}\frac{-\omega Im\Pi_{\mu}(\mathbf{q},\omega)}{\left[\omega^{2}-\tilde{\Omega}_{
 \mu}^{2} (\mathbf{q})-Re\Pi_{\mu}(\mathbf{q},\omega)\right]^{2}+ \left[Im\Pi_{\mu}(\mathbf{q},\omega)\right]^{2}}.
\label{ins-eq}
\end{multline}
Peaks in $\sigma(\mathbf{q},\omega)$ represent phonon excitations observed experimentally. 
Replacing $\Pi_{\mu}(\mathbf{q},\omega) \to \Pi_{\mu}(\mathbf{q},\tilde{\Omega}_{\mu}(\mathbf{q}))$ in Eq. \eqref{ins-eq}
the Lorentzian approximation is recovered, in which each peak is represented 
with a Lorentzian function centered at 
$
 \Omega_{\mu}(\mathbf{q})=\sqrt{\tilde{\Omega}_{\mu}^{2}(\mathbf{q})+Re\Pi_{\mu}(\mathbf{q},\tilde{\Omega}_{\mu}(\mathbf{q}))}
 \label{bubble}
$
with a linewidth proportional to $Im\Pi_{\mu}(\mathbf{q},\tilde{\Omega}_{\mu}(\mathbf{q}))$\cite{PhysRevB.97.214101}.
  
\begin{figure*}[t]
\includegraphics[width=0.43\linewidth]{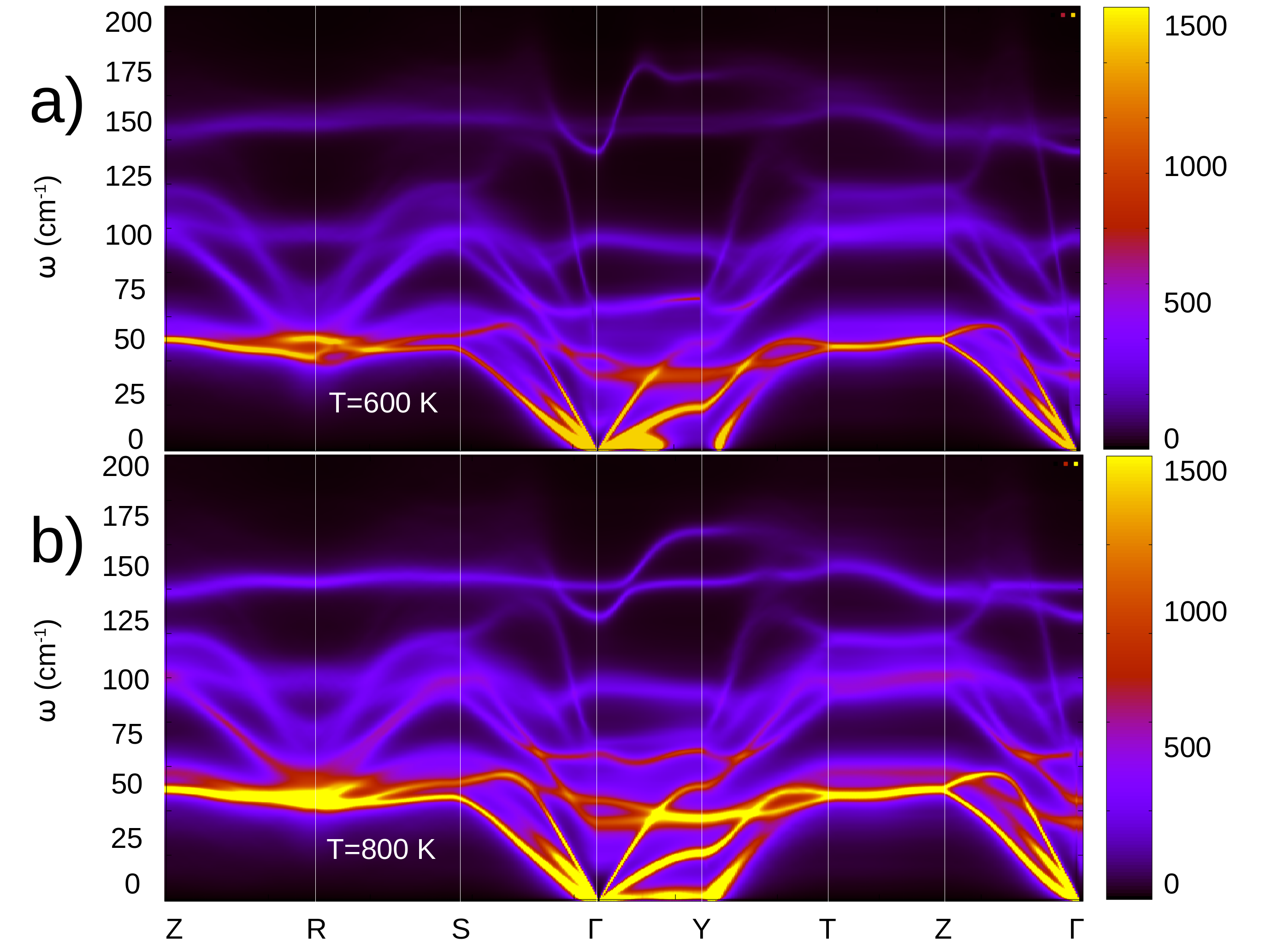}
\includegraphics[width=0.53\linewidth]{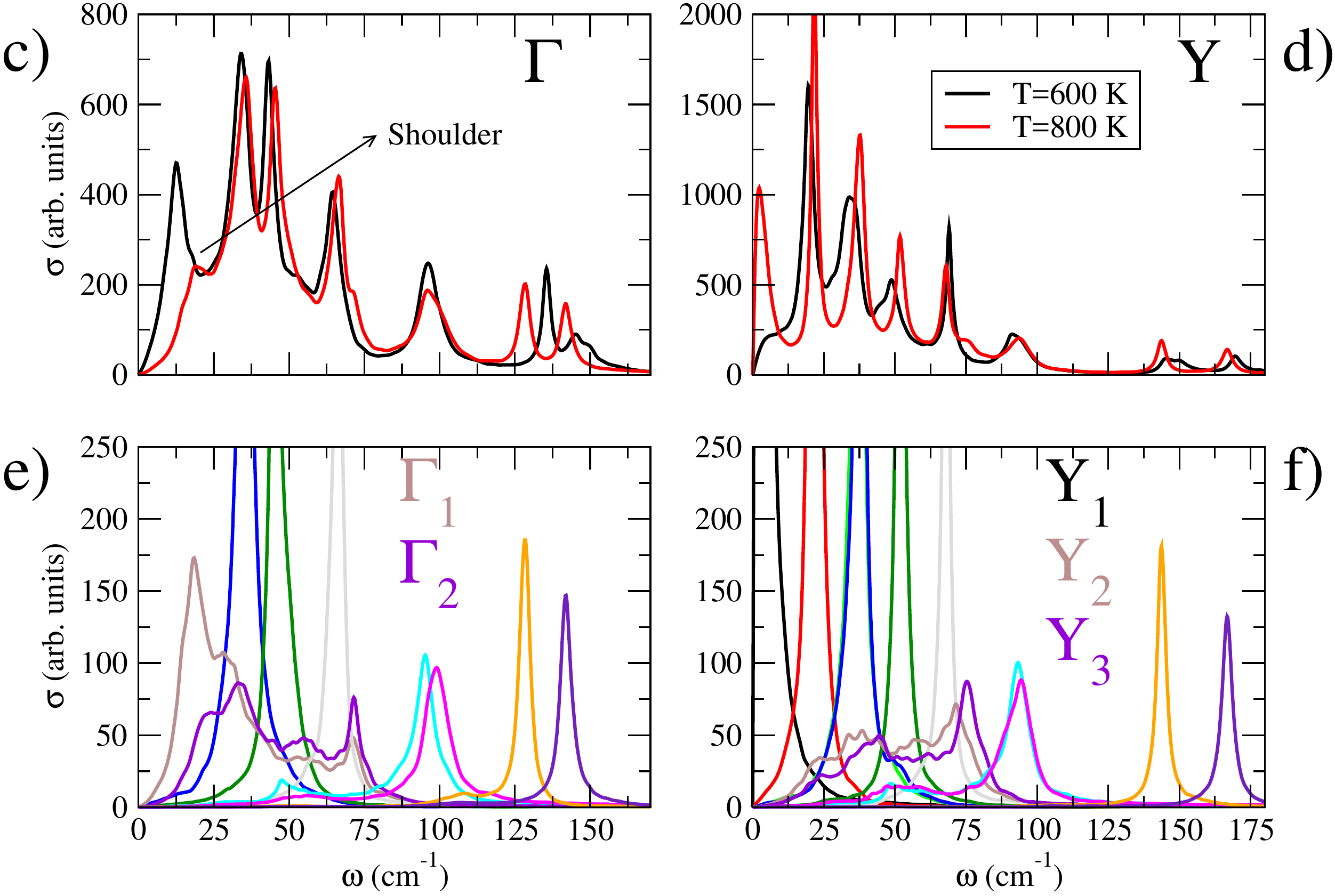}
\caption{Spectral function of SnSe in the $Cmcm$ phase calculated at a) $600$ K and b) 
$800$ K using $\overset{(3)}{\mathbf{\Phi}}$ at the correponding temperature. 
The spectral function at the c) $\Gamma$ and d) Y points at $600$ and $800$ K. The 
contribution of each mode to the spectral function is also shown at the $\Gamma$ point e) and the Y point f) at $800$ K. 
Different colors correspond to different modes. All the calculations are performed within LDA in the experimental structure. 
In each case we use $\tilde{\Omega}_{\mu}(\mathbf{q})$ calculated at the same temperature as $\overset{(3)}{\mathbf{\Phi}}$.}
\label{ins}
\end{figure*}

Fig. \ref{spectrum} compares the harmonic phonon spectrum with the 
anharmonic one in the Lorentzian approximation obtained at $800$ K within LDA in the experimental lattice (the results below are also obtained
within the LDA in the experimental lattice). 
The  anharmonic correction is very large for most of the modes across the Brillouin zone. 
Within the harmonic approximation, there are five unstable modes: 
two ($\Gamma_{1}$, $\Gamma_{2}$) at $\Gamma$, two ($Y_{1}$, $Y_{3}$) at $Y$ and one ($R_{1}$) at $R$. 
The instabilities at $\Gamma$ would cause ferroeletric  
transitions\cite{hong2016electronic,skelton2016anharmonicity}, but
they suffer from a huge anharmonic renormalization that prevents it. $Y_{3}$ and $R_{1}$ are also stabilized by 
anharmonic effects. The $Y_1$ mode however, even if it is strongly affected by anharmonicity,
remains unstable at $600$ K (see Fig. \ref{ins}a), i.e., the $Cmcm$ phase is not a minimum of the free energy
and the crystal distorts adopting the low-symmetry $Pnma$ phase. 

In highly anharmonic materials\cite{delaire2011giant,Li2014phonon,
ribeiro2017anharmonic,ribeiro2017anharmonic,Li2014phonon,paulatto2015first,PhysRevB.97.214101}, the 
spectral functions show broad peaks, shoulders and satellite peaks,
strongly deviating from the Lorentzian picture.
In Fig. \ref{ins} we show the spectral function 
keeping the full frequency dependence on the self-energy, without assuming the Lorentzian
lineshape. The spectral function clearly reproduces the collapse of the $Y_1$ mode at the 
transition temperature. 
The calculated spectral functions show that the strong anharmonicity
present on the phonon frequency renormalization is also reflected on the
spectral function.
The strongly anharmonic features specially affect 
in-plane modes in the 25-75 cm$^{-1}$
energy range. 
For instance, at the $\Gamma$ point the
$\Gamma_1$ mode, who describes a vibration along the in-plane $z$ axis in opposite
direction for the Sn and Se atoms (see Fig. \ref{structure}) and is stabilized by anharmonicity,
shows a double peak structure and a broad shoulder (see Fig. \ref{ins}e). 
The mode that describes the same vibration ($\Gamma_2$) but in the
other in-plane $y$ direction also shows a very complex non-lorentzian shape.
The overall $\sigma(\mathbf{q}=\Gamma,\omega)$ 
consequently has a broad shoulder at $\simeq$ 25 cm$^{-1}$ as marked 
in Figs.  \ref{ins}c, which is less acute as temperature increases. 
At the Y point there are also two modes, $Y_{2}$,
whose eigenvector is plotted in Fig. \ref{structure}, and $Y_{3}$,
which describes the same displacement but in the other $z$ in-plane 
direction, that show a strongly anharmonic non-Lorentzian shape.   
The modes with very complex line-shapes are those that show the largest
linewidth in the Lorentzian limit (see Fig. \ref{spectrum}), 
for instance, $12$ cm$^{-1}$ for the half width at half maximum of the $\Gamma_{1}$ mode.
These modes have strongly anomalous spectral functions
and large linewidths because they can easily scatter
with an optical mode close in energy and an acoustic mode close to $\Gamma$.
We identify this by directly analyzing which phonon triplets contribute more to the linewidth.
It is interesting to remark that if $\Pi_{\mu}(\mathbf{q},\omega)$
is calculated by substituting $\overset{(3)}{\boldsymbol\Phi}$ by $\overset{(3)}{\boldsymbol\phi}$, the 
anomalies of these modes become weaker (see Supplementary Material). 
This underlines that in the $Cmcm$ phase the third-order derivatives of $V$ 
are not sufficient to calculate the phonon linewidths and
that higher order terms are important, which are effectively captured by    
$\overset{(3)}{\mathbf{\Phi}}$.

In Fig. \ref{experiments} we present the lattice thermal conductivity
calculated with the SSCHA frequencies ($\tilde{\Omega}_{\mu}(\mathbf{q})$) and 
non-perturbative TOFCs ($\overset{(3)}{\boldsymbol{\Phi}}$). For comparison we also 
calculate $\kappa_l$ substituting $\overset{(3)}{\boldsymbol{\Phi}}$ by $\overset{(3)}{\boldsymbol{\phi}}$.
The calculation is performed solving the BTE assuming the single-mode relaxation time approximation (SMA), whose 
validity was confirmed against a more accurate iterative method \cite{fugallo2013ab} (see Supplementary Material).
The thermal conductivity of SnSe is very low, mainly because
the contribution of optical modes is strongly suppressed by the large 
anharmonicity and because the contribution of acoustic modes
is also reduced due to the large scattering among themselves and with the $\Gamma_1$ 
mode.  
We compare these results with the values obtained by 
Ibrahim et al.\cite{ibrahim2017reinvestigation} above the possible transition at 600 K 
(only the in-plane $\kappa_l$ are reported at these temperatures)
and with the values obtained by Zhao et al.\cite{zhao2014ultralow} above the transition at 800 K.
The lattice thermal conductivity is in better agreement
with experimental results using $\overset{(3)}{\boldsymbol{\Phi}}$ instead of $\overset{(3)}{\boldsymbol{\phi}}$, 
which overestimates the lattice thermal conductivity along the in-plane $y$ and $z$ directions. 
This is consistent with the larger phonon linewidths obtained with the 
non-perturbative TOFCs. The agreement for the in-plane $\kappa_{yy} \sim \kappa_{zz}$
with the measurements by Ibrahim et al.\cite{ibrahim2017reinvestigation} is good 
in the non-perturbative limit, contrary to previous calculations that
underestimate it\cite{skelton2016anharmonicity}. The calculated out-of-plane $\kappa_{xx}$ 
is also in good agreement with the results by Zhao et al.\cite{zhao2014ultralow},
but we find that their ultralow results for the in-plane $\kappa_l$, in contradiction
with the values in Ref. \onlinecite{ibrahim2017reinvestigation}, are underestimated.
It is not surprising that the thermal conductivity in vacancy free SnSe 
is lower along the out-of-plane direction due to the weaker bonding. 
Thus, our calculations support the interpretation that the weak anisotropy and ultralow
thermal conductivity measured by Zhao et al.\cite{zhao2014ultralow}
above the transition was produced by the large amount of Sn vacancies 
present in their samples\cite{ibrahim2017reinvestigation,Wei2016,Zhao2016}.

In conclusion, our first-principles calculations
show that the vibrational properties of SnSe in the $Cmcm$ phase
are dominated by huge non-perturbative anharmonic effects. 
We show how the collapse of the 
$Y_1$ mode is responsible for the second-order phase transition 
between the high-symmetry $Cmcm$ and the low-symmetry $Pnma$ phase.  
The calculated transition temperature is volume and 
functional dependent.
The spectral functions of in-plane modes are characterized
by very anomalous features, with shoulders and double peaks, clearly
deviating from the standard Lorentzian-like shape.
These results will eventually be crucial to interpret future INS experiments
for the high-temperature phase.
The calculated in-plane thermal conductivity is in good agreement with the
experiments by Ibrahim et al.\cite{ibrahim2017reinvestigation}, but
not with those by Zhao et al.\cite{zhao2014ultralow}, supporting
the interpretation that in the latter experiment the 
thermal conductivity was lowered by Sn vacancies.
Our results show for the first time that 
the inclusion of non-perturbative effects
is crucial not only for renormalizing phonon spectra, but also
for obtaining third-order force-constants that yield
a lattice thermal conductivity in agreement with experiments.
Similar huge non-perturbative anharmonic effects are expected in other
good thermoelectric materials. 

The authors acknowledge fruitful discussions with O. Delaire. 
Financial support was provided by the 
Spanish Ministry of Economy and Competitiveness (FIS2013- 48286-C2-2-P), 
the Department of Education, Universities and Research of the Basque Government 
and the University of the Basque Country (IT756-13).
U.A. is also thankful to the Material Physics Center for a predoctoral fellowship. Computer 
facilities were provided by the 
Donostia International Physics Center (DIPC), the Spanish Supercomputing Network 
(FI-2017-2-0007) and PRACE (2017174186).

\end{document}


\maketitle

\section{Structure and high symmetry points}

The $Cmcm$ and $Pnma$ phases are orthorhombic and their structure  is shown in the main text. The $Cmcm$ phase has a centering in the $XY$ plane 
and contains 4 atoms in the primitive cell. The primitive cell of the $Pnma$ phase contains $8$ atoms. The primitive lattice vectors of the $Cmcm$ 
structure are: $\mathbf{a}_{1}=(a/2,b/2,0)$, $\mathbf{a}_{2}=(-a/2,b/2,0)$ and $\mathbf{a}_{3}=(0,0,c)$, where $a$, $b$ and $c$ are the lattice 
constants of the rectangular primitive cell of the $Pnma$ phase. The reciprocal lattice vectors of the $Cmcm$ primitive first Brillouin zone are 
$\mathbf{b}_{1}=2\pi(1/a,1/b,0)$, $\mathbf{b}_{1}=2\pi(-1/a,1/b,0)$ and $\mathbf{b}_{1}=2\pi(0,0,1/c)$.
The high symmetry points and their coordinates used in phonon dispersion figures are listed in table \ref{qpoints}.
\begin{table}
\begin{center}
\begin{tabular*}{0.45\textwidth}{l c}
 \hline
 \hline
             Symmetry point  & Reduced $\mathbf{q}$ vector  \\
 \hline
 $Z$                  &  0.0, 0.0, 0.5  \\
 $R$                  &  0.0, 0.5, 0.5  \\
 $S$                  &  0.0, 0.5, 0.0  \\
 $\Gamma$        &  0.0, 0.0, 0.0  \\
 $Y$                  &  0.5, 0.5, 0.0  \\
 $T$                  &  0.5, 0.5, 0.5  \\
 \hline
 \hline
\end{tabular*}
\end{center}
\caption{Reduced $\mathbf{q}$ vectors of the high symmetry points in the Brillouin zone of the $Cmcm$ phase. The coordinates are given with 
respect to the reciprocal lattice vectors of the primitive cell.}
\label{qpoints}
\end{table}

\section{Calculation methods, pseudopotentials and parameters}

\begin{figure}[th]
\begin{center}
\includegraphics[width=\linewidth]{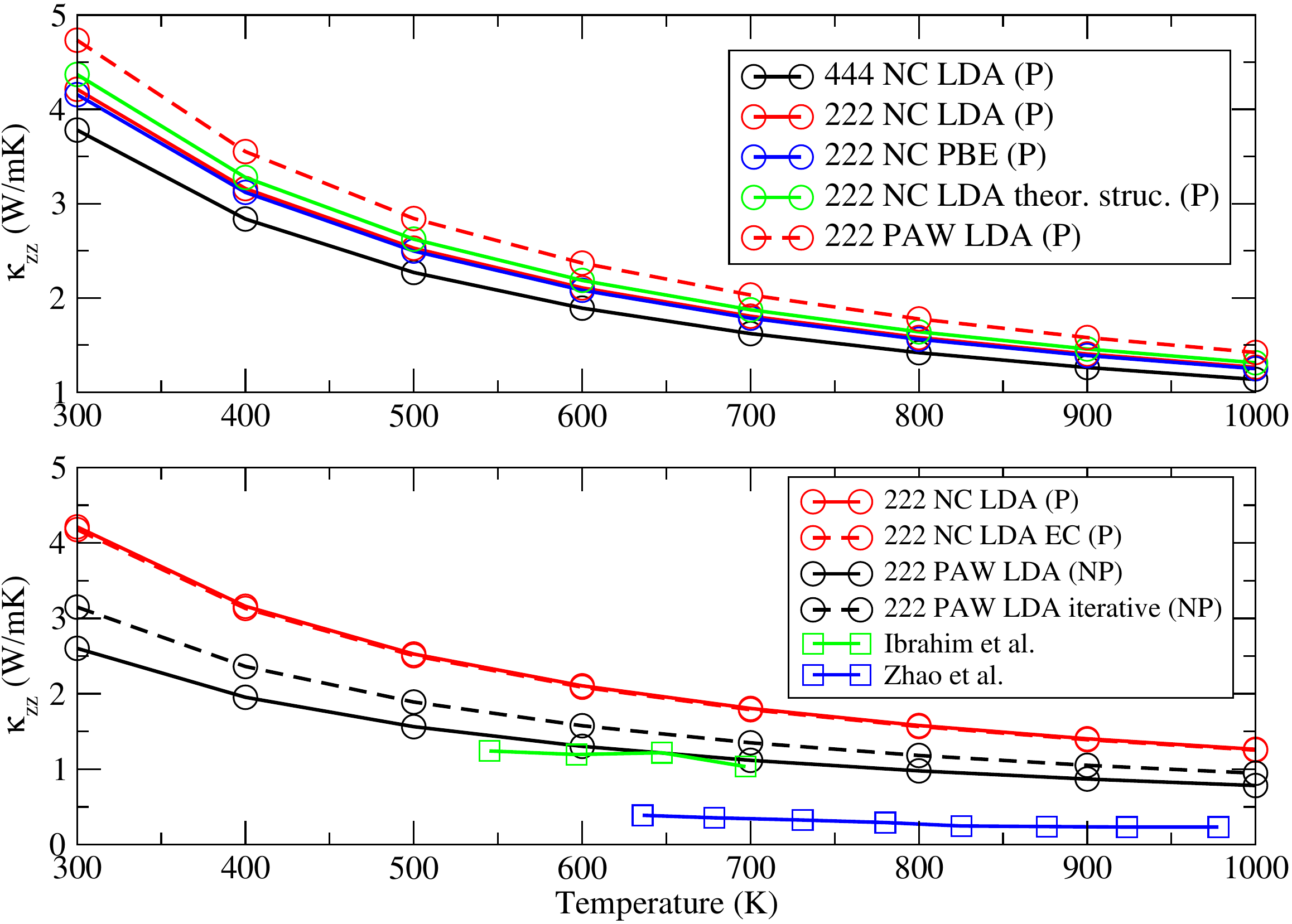}
\caption{$\kappa_{zz}$ component of the lattice thermal conductivity within different approximations (Perturbative (P) and non-perturbative (NP)), 
parameters, structures and pseudopotentials (see legends). 
a) Thermal conductivity of SnSe using norm-conserving (NC) LDA pseudopotential and a $2\times2\times2$ supercell (solid red), 
NC LDA pseudopotential and a $4\times4\times4$ supercell (black), PAW LDA pseudopotential with a $2\times2\times2$ supercell (dashed red), 
NC PBE pseudopotential with a $2\times2\times2$ supercell (solid blue) and NC LDA pseudopotential with a $2\times2\times2$ supercell in 
the fully relaxed structure (solid green). b) Thermal conductivity of SnSe using NC LDA pseudopotential and a $2\times2\times2$ supercell (solid 
red), NC LDA pseudopotential including effective charges (EC) in a $2\times2\times2$ supercell (dashed red), non-perturbative (NP) PAW LDA 
calculation using a $2\times2\times2$ supercell and PAW LDA calculation using the iterative method in a $2\times2\times2$ supercell. Experimental 
points are shown as squares. All calculations are done using the experimental structure but the green data in a). In the experimental lattice we 
use the $\tilde{\Omega}_{\mu}$ phonons at $700$ K in LDA and $500$ K in PBE. For the calculation in the theoretical 
structure we use the $\tilde{\Omega}_{\mu}$ phonons at $400$ K. The other Cartesian components show a similar behavior.}
\label{tk-ps-sc}
\end{center}
\end{figure}

Our calculations are based on Density Functional Theory (DFT) using the QUANTUM-ESPRESSO\cite{giannozzi2009quantum} software 
package. Harmonic phonons were calculated within Density Functional Perturbation Theory\cite{baroni2001phonons} and perturbative third-order 
force-constants (TOFC) were calculated within Density Functional Pertubation Theory\cite{paulatto2013anharmonic} (DFPT) or finite 
differences\cite{ShengBTE_2014}. Anharmonic phonons and non-perturbative TOFCs were calculated using the SSCHA. For the exchange-correlation 
interaction we use the Perdew-Burke-Ernzerhof\cite{perdew1996generalized} (PBE) generalized gradient approximation and the local density 
approximation (LDA) with ultrasoft\cite{kresse1994norm} (US) and Projector Augmented Wave method\cite{blochl1994projector} (PAW) pseudopotentials 
respectively. Due to limitations in the implementation of perturbative TOFCs within DFPT we use norm-conserving (NC) 
pseudopotentials\cite{troullier1991efficient} and finite differences within PBE. In the LDA case we can see the comparison with the PAW 
pseudopotential (using a finite difference approach\cite{li2014shengbte}) for the thermal conductivity in Fig. \ref{tk-ps-sc}. We also 
include the calculation with effective charges in the phonon spectrum, as we can see, the difference is negligible. The calculation beyond the 
Single Mode Approximation (SMA) (the iterative method\cite{fugallo2013ab}) is also included. As it is expected for high temperatures, the effect is 
small. The thermal conductivity in the fully relaxed structure is also included, as we can see the volume effect is small. We use a cutoff energy 
of $70$ Ry and a grid of $16\times16\times16$ $\mathbf{K}$ points to sample the first Brillouin zone. For the harmonic phonon calculations we use a 
$6\times6\times6$ supercell and a $2\times2\times2$ one for the TOFCs. The convergence of the TOFCs is tested in the thermal conductivity with a 
$4\times4\times4$ supercell in Fig. \ref{tk-ps-sc} and the difference is small. For the SSCHA calculation we use a $2\times2\times2$ 
supercell. Then, for the phonon dispersion, we interpolate the difference between the harmonic and anharmonic force-constants in the 
$2\times2\times2$ supercell to the $6\times6\times6$ and we add it to the $6\times6\times6$ harmonic force constants. For the linewidth 
calculations we use a $20\times20\times20$ $\mathbf{q}$ points grid obtained by Fourier interpolation with a smearing of $1$ cm$^{-1}$. For the 
thermal conductivity calculation we use a $10\times10\times10$ grid of $\mathbf{q}$ points to sample the first Brillouin zone. We have done 
calculations in denser grids to test these parameters.

\section{Volume effects in the harmonic phonon spectrum}

\begin{figure}[th]
\begin{center}
\includegraphics[width=0.8\linewidth]{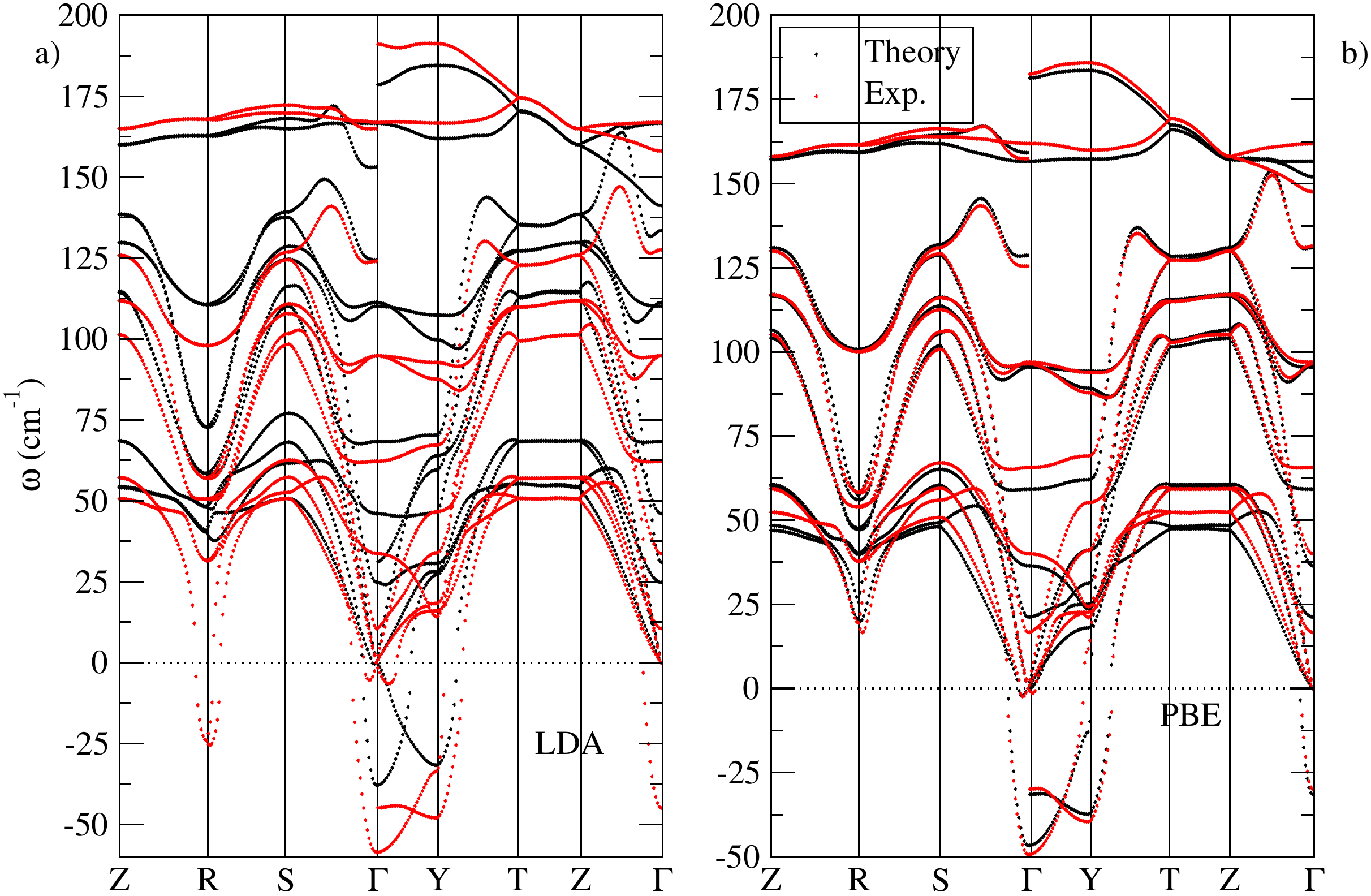}
\caption{a) LDA harmonic phonon spectra in the theoretical and experimental structures. b) The same as a) within PBE. The LO-TO splitting is 
included in the calculations.}
\label{phonon}
\end{center}
\end{figure}

In this section we will analyze the volume effects in the harmonic phonon spectrum of $Cmcm$ SnSe. In Fig. \ref{phonon} we can see the LDA and 
PBE phonon spectra in the theoretical and experimental structures.
As we can see, within LDA, the difference between phonons in the theoretical and experimental structures is clearly visible. Due to the bigger 
volume in the experimental cell, there is a red shift of almost all the vibrational modes and furthermore, more imaginary frequencies arise at the 
$\Gamma$ and $Y$ points. Within PBE, the phonon spectra in the theoretical and experimental structures are much more similar, because the lattice 
parameters in these structures are very similar as well. These results are in agreement with our $T_{c}$ calculations shown in the main text, where 
$T_{c}$ is basically the same in the two structures within PBE and completely different within LDA. We conclude that the harmonic phonons of SnSe 
are very sensitive to the volume of the unit cell.

\begin{figure}[th]
\begin{center}
\includegraphics[width=0.5\linewidth]{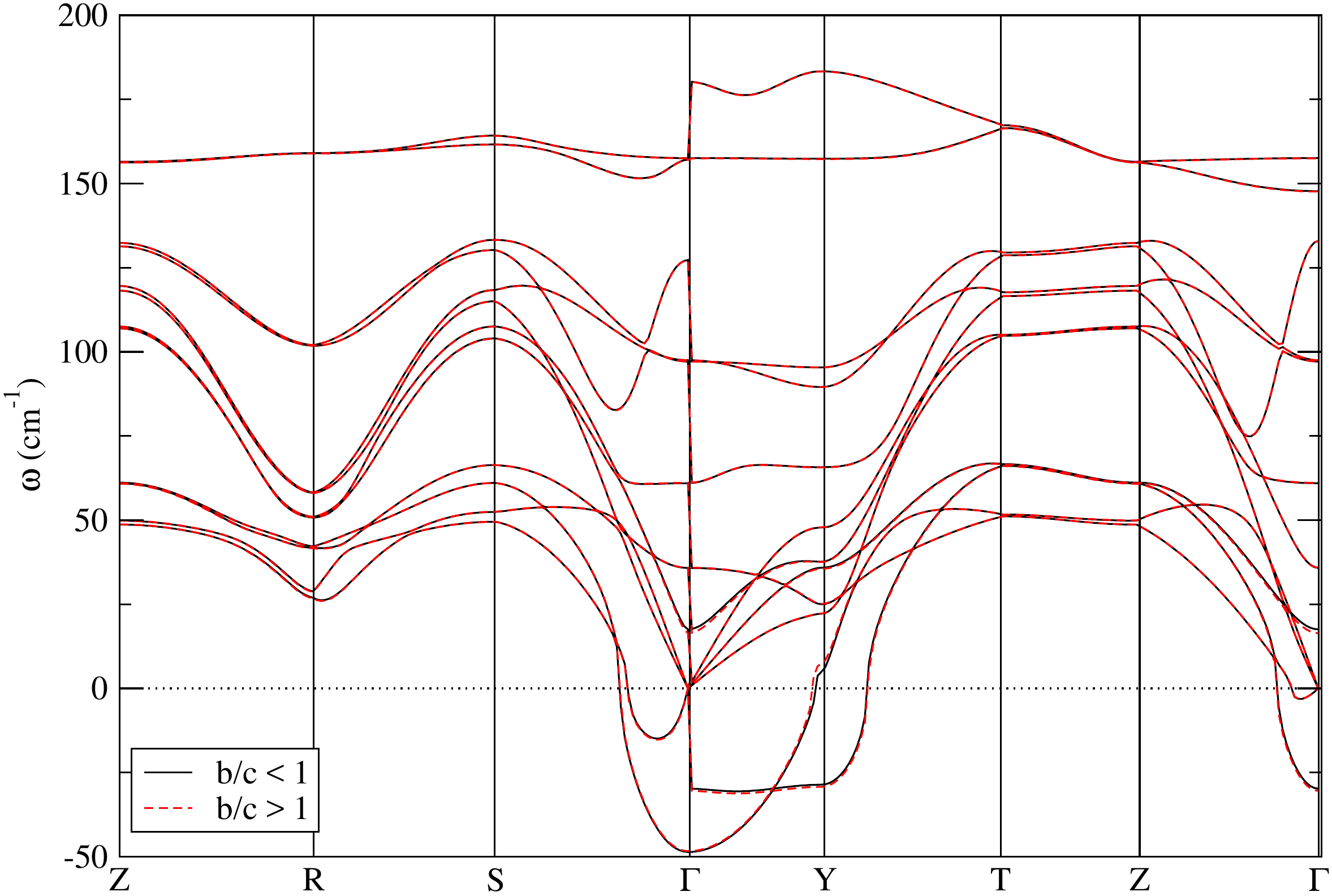}
\caption{PBE harmonic phonon spectra with the following lattice parameters: $a=22.492$, $b=8.094$ and $c=8.090$ (in Bohr units) (red) in one 
case and exchanging $b$ and $c$ (black) in the other. We have done the calculation in a $2\times2\times2$ supercell and Fourier
interpolated to plot the spectrum. The LO-TO splitting is included in the calculations.}
\label{lattice}
\end{center}
\end{figure}

In order to compare with the calculations in Ref. \cite{dewandre2016two} and see whether the change from $b/c<1$ to $b/c>1$ is responsible 
for creating the instability at the point $Y$, we have done the phonon calculation using the lattice parameters of Ref. \cite{dewandre2016two}. The 
result is shown in Fig. \ref{lattice}.
It is clear that the change in the lattice parameters does not alter the instability, as expected for such a small change.

\section{Anharmonic theory: SSCHA}

We study the lattice dynamics of SnSe in the Born-Oppenheimer (BO) approximation, thus we consider the quantum Hamiltonian for the atoms defined by 
the BO potential energy $V(\mathbf{R})$. With $\mathbf{R}$ we are denoting in component-free notation the quantity $R^{s\alpha}(\mathbf{l})$, which 
is a collective coordinate that completely specifies the atomic configuration of the crystal. The index $\alpha$ denotes the Cartesian direction, 
$s$ labels the atom within the unit cell and $\mathbf{l}$ indicates the three dimensional lattice vector. In what follows we will also use a single 
composite index $a=(\alpha,s,\mathbf{l})$ to indicate Cartesian index $\alpha$, atom $s$ index and lattice vector $\mathbf{l}$. Moreover, in 
general, we will use bold letters to indicate also other quantities in component-free notation.

In order to take into account quantum effects and anharmonicity at a non-perturbative level, we use the Stochastic Self-Consistent Harmonic 
Approximation (SSCHA). For a given temperature $T$, the method allows to find an approximation for $F(\boldsymbol{\mathcal{R}})$, the free energy 
of the crystal as a function of the average atomic position $\mathcal{R}^{a}$ (the centroids). For a given centroid $\boldsymbol{\mathcal{R}}$, the 
SSCHA free energy is obtained through an auxiliary quadratic Hamiltonian, the SSCHA Hamiltonian $\mathcal{H}_{\boldsymbol{\mathcal{R}}}$. In a 
displacive second-order phase transition, at high temperature the free energy has a minimum in a high symmetry configuration 
$\boldsymbol{\mathcal{R}}_{hs}$ but, on lowering temperature, $\boldsymbol{\mathcal{R}}_{hs}$ becomes a saddle point at the transition temperature 
$T_{c}$. Therefore, the free energy Hessian evaluated in $\boldsymbol{\mathcal{R}}_{hs}$, 
$\partial^{2}F/\partial\boldsymbol{\mathcal{R}}\partial\boldsymbol{\mathcal{R}}|_{\boldsymbol{\mathcal{R}}_{hs}}$, 
at high temperature is positive definite but it develops one or multiple negative eigendirections at $T_{c}$. The SSCHA free energy Hessian can be 
computed by using the analytic formula (in component-free notation)
\begin{equation}\label{formula}
 \frac{\partial^{2}F}{\partial\boldsymbol{\mathcal{R}}\partial\boldsymbol{\mathcal{R}}}=\mathbf{\Phi} + \overset{(3)}{\mathbf{\Phi}}\mathbf{
 \Lambda}(0)[\mathbf{1}-\overset{(4)}{\mathbf{\Phi}}\mathbf{\Lambda}(0)]^{-1}\overset{(3)}{\mathbf{\Phi}},
\end{equation}
with
\begin{equation}
 \mathbf{\Phi}=\left\langle\frac{\partial^{2}V}{\partial\mathbf{R}\partial\mathbf{R}}\right\rangle_{\rho_{\mathcal{H}_{\mathbf{\mathcal{R
 }}}}}, \overset{(3)}{\mathbf{\Phi}}=\left\langle\frac{\partial^{3}V}{\partial\mathbf{R}\partial\mathbf{R}\partial{\mathbf{R}}}\right\rangle_{
 \rho_{\mathcal{H}_{\mathbf{\mathcal{R}}}}}, \overset{(4)}{\mathbf{\Phi}}=\left\langle\frac{\partial^{4}V}{\partial\mathbf{R}\partial\mathbf{
 R}\partial\mathbf{R}\partial\mathbf{R}}\right\rangle_{\rho_{\mathcal{H}_{\mathbf{\mathcal{R}}}}}, 
\end{equation}
where the averages are with respect to the density matrix of the SSCHA Hamiltonian $\mathcal{H}_{\mathbf{\mathcal{R}}}$. i.e $\rho_{
\mathcal{H}_{\mathbf{\mathcal{R}}}}=e^{-\beta\mathcal{H}_{\mathbf{\mathcal{R}}}}/tr[e^{-\beta\mathcal{H}_{\mathbf{\mathcal{R}}}}]$
(labeled as $\rho_{\mathbf{\mathcal{R}},\mathbf{\Phi}}$ in the main text), and 
$\beta=(K_{B}T)^{-1}$ where $K_{B}$ is the Boltzmann constant. In Eq. \ref{formula} the value $z=0$ of the 4th-order tensor $\mathbf{\Lambda}(z)$ 
is used. For a generic complex number $z$ it is defined, in components, by
\begin{equation}
 \mathbf{\Lambda}^{abcd}(z)=-\frac{1}{2}\sum_{\mu\nu}\tilde{F}(z,\tilde{\Omega}_{\mu},\tilde{\Omega}_{\nu})\times \sqrt{\frac{\hbar}{2M_{a}\tilde{
 \Omega}_{\mu}}}e_{\mu}^{a}\sqrt{\frac{\hbar}{2M_{b}\tilde{\Omega}_{\nu}}}e_{\nu}^{b}\sqrt{\frac{\hbar}{2M_{c}\tilde{\Omega}_{\mu}}}e_{\mu}^{
 c}\sqrt{\frac{\hbar}{2M_{d}\tilde{\Omega}_{\nu}}}e_{\nu}^{d},
\end{equation}
with $M_{a}$ the mass of the atom $a$, $\tilde{\Omega}_{\mu}^{2}$ and $e_{\mu}^{a}$ eigenvalues and corresponding eigenvectors of $D_{ab}^{(S)}=
\mathbf{\Phi}_{ab}/\sqrt{M_{a}M_{b}}$, respectively, and 
\begin{equation}
 \tilde{F}(z,\tilde{\Omega}_{\mu},\tilde{\Omega}_{\nu})=\frac{2}{\hbar}\left[\frac{(\tilde{\Omega}_{\mu}+\tilde{\Omega}_{\nu})[1+n_{B}(\tilde{
 \Omega}_{\mu})+n_{B}(\tilde{\Omega}_{\nu})]}{(\tilde{\Omega}_{\mu}+\tilde{\Omega}_{\nu})^{2}-z^{2}} -\frac{(\tilde{\Omega}_{\mu}-\tilde{
 \Omega}_{\nu})[n_{B}(\tilde{\Omega}_{\mu})-n_{B}(\tilde{\Omega}_{\nu})]}{(\tilde{\Omega}_{\mu}-\tilde{\Omega}_{\nu})^{2}-z^{2}}\right]
\end{equation}
where $n_{B}(\omega)=1/(e^{\beta\hbar\omega}-1)$ is the bosonic occupation number. Evaluating through Eq. \ref{formula} the free energy Hessian in 
$\boldsymbol{\mathcal{R}}_{hs}$ and studying its spectrum as a function of temperature, we can predict the occurrence of a displacive second-order 
phase transition and estimate $T_{c}$. In particular, since we are considering a crystal, we take advantage of lattice periodicity 
and we Fourier transform the free energy Hessian with respect to the lattice indexes. Therefore, since there are $4$ atoms in the unit cell of 
SnSe, we actually calculate the eigenvalues of the two dimensional $12\times12$ square matrix $\partial^{2}F/
\partial\mathcal{R}^{a}(-\mathbf{q})\partial\mathcal{R}^{b}(\mathbf{q})$ in different $\mathbf{q}$ of the Brillouin zone. 

\begin{figure}[th]
\begin{center}
\includegraphics[width=0.45\linewidth]{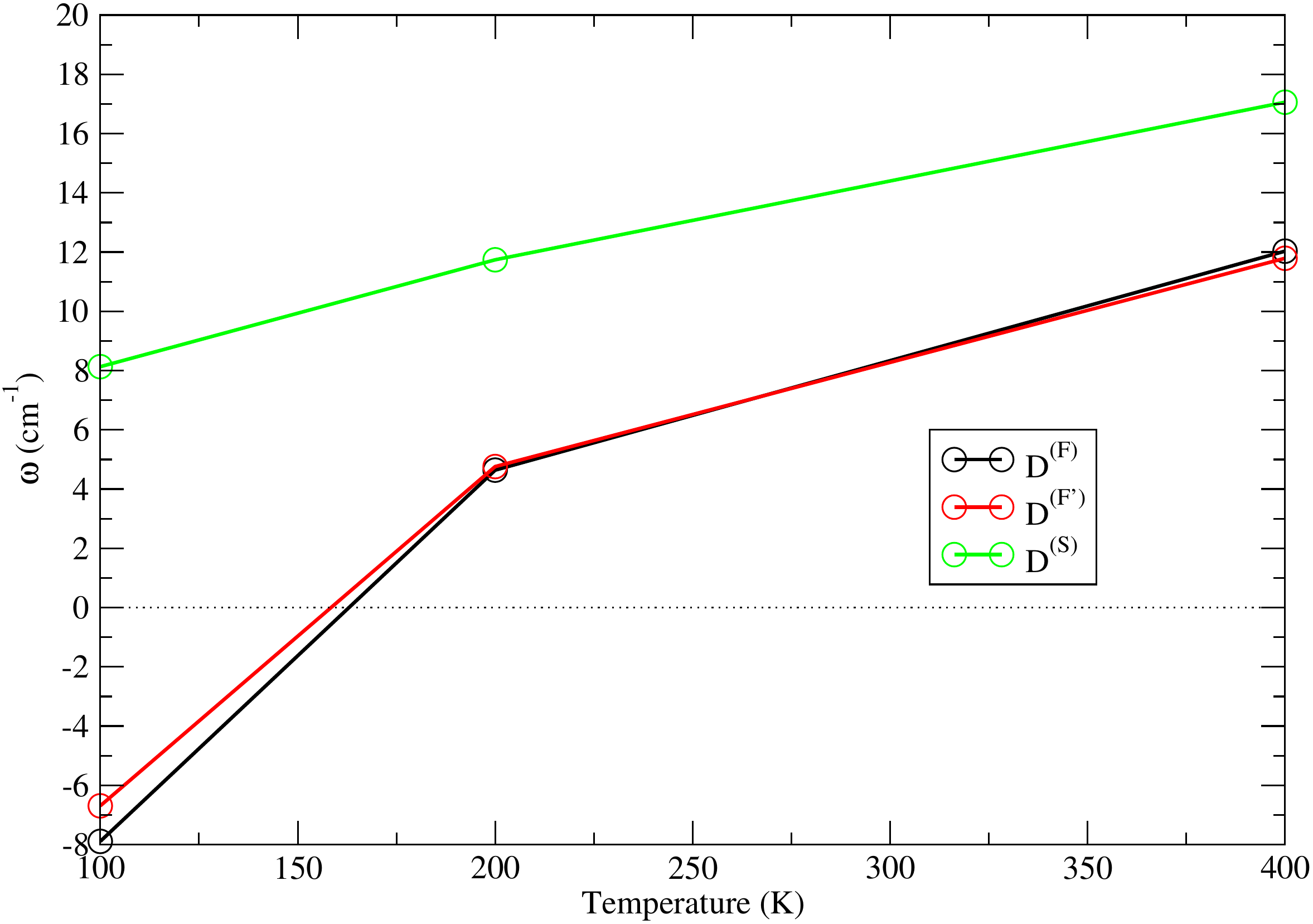}
\includegraphics[width=0.45\linewidth]{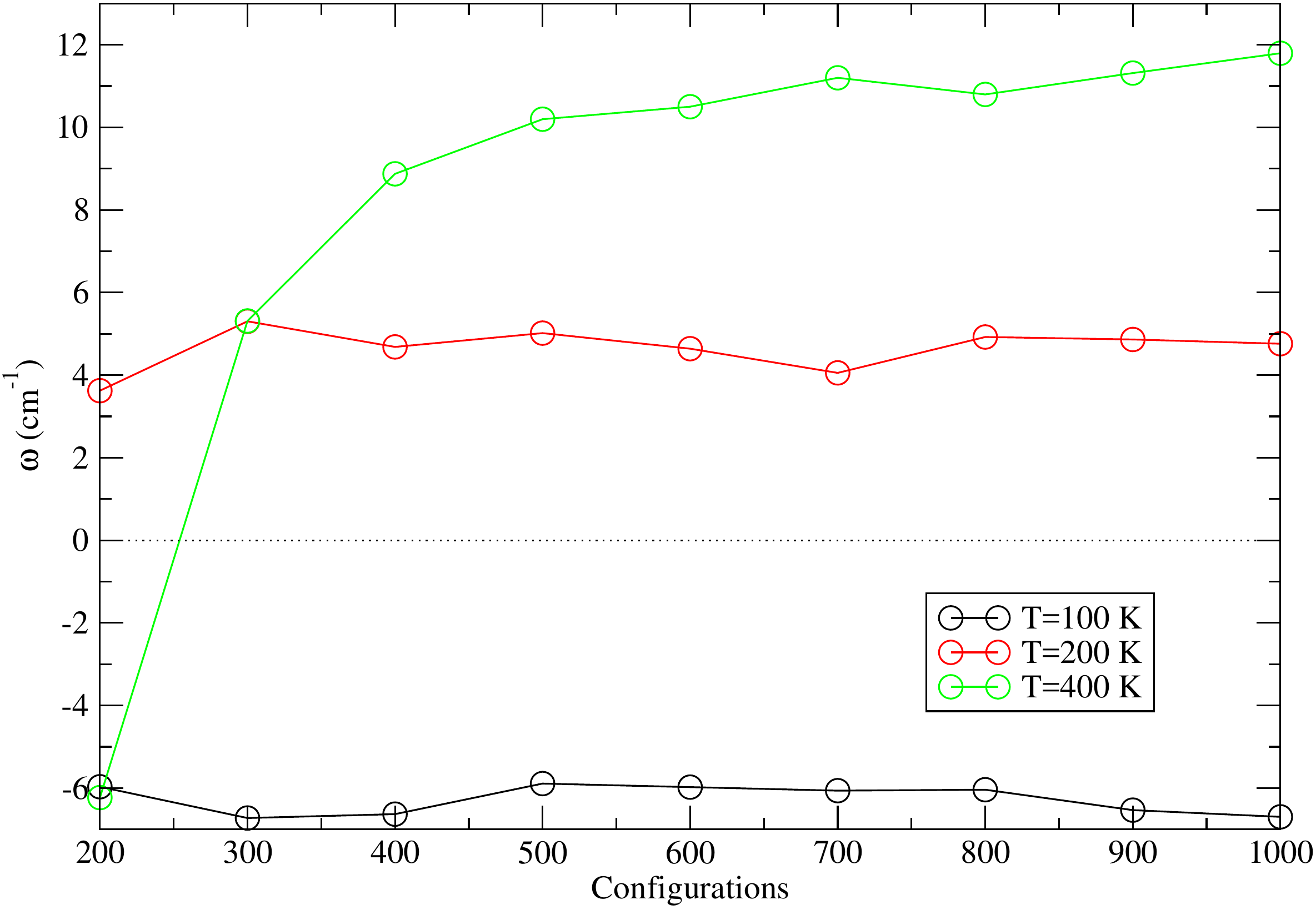}
\caption{Left panel: Frequency of the mode $Y_{1}$ as a function of temperature within LDA in the theoretical structure within different 
approximations. The black circles corresponds to $\mathbf{D}^{(F)}$, the green ones to $\mathbf{D}^{(S)}$ and the red ones to $\mathbf{D}^{(F')}$, 
which is the same as $\mathbf{D}^{(F)}$ with $\overset{(4)}{\mathbf{\Phi}}=0$. Right panel: We show the convergence of
the frequency calculated from $D^{(F')}$ with respect to the number of stochastic configurations used for the SSCHA at different temperatures. The 
convergence is the same for $D^{(F)}$.}
\label{odd3}
\end{center}
\end{figure}

As shown in ref. \cite{bianco2017second}, in the context of the SSCHA it is possible to formulate an ansatz in order to give an approximate 
expression to the one-phonon Green function $\mathbf{G}(z)$ for the variable $\sqrt{M_{a}}(R^{a}-\mathcal{R}_{hs}^{a})$:
\begin{equation} \label{green}
 \mathbf{G}^{-1}(z)=z^{2}\mathbf{1}-\mathbf{M}^{-\frac{1}{2}}\mathbf{\Phi}\mathbf{M}^{-\frac{1}{2}}- \mathbf{\Pi}(z),
\end{equation}
where $\mathbf{G}^{-1}(0)=-\mathbf{D}^{(F)}$, $D_{ab}^{(F)}=\frac{1}{\sqrt{M_{a}M_{b}}}\frac{\partial^{2}F}{\partial\mathbf{\mathcal{R}}_{a}
\partial\mathbf{\mathcal{R}}_{b}}$, and $\Pi(z)$ is the SSCHA self-energy, given by
\begin{equation}
 \Pi(z)=\mathbf{M}^{-\frac{1}{2}}\overset{(3)}{\mathbf{\Phi}}\mathbf{\Lambda}(z)[\mathbf{1}-\overset{(4)}{\mathbf{\Phi}}\mathbf{\Lambda}(z)]^{-1}
 \overset{(3)}{\mathbf{\Phi}}\mathbf{M}^{-\frac{1}{2}},
\end{equation}
where $M_{ab}=\delta_{ab}M_{a}$ is the mass matrix. As can be seen in Fig. \ref{odd3}, for the applications considered in the 
present paper, the static term $\overset{(4)}{\mathbf{\Phi}}\mathbf{\Lambda}(0)$ is negligible  
with respect to the identity matrix. We can also see the convergence of the frequencies with respect to the amount of stochastic configurations.
Extending this approximation to the dynamical case reduces the SSCHA self-energy to the so called bubble self-energy, namely
\begin{equation}
 \mathbf{\Pi}\approx\mathbf{\Pi}^{(B)}(z)=\mathbf{M}^{-\frac{1}{2}}\overset{(3)}{\mathbf{\Phi}}\mathbf{\Lambda}(z)\overset{(3)}{\mathbf{
 \Phi}}\mathbf{M}^{-\frac{1}{2}}.
\end{equation}
We then neglect the mixing between different phonon modes and assume that $\mathbf{\Pi}(z)$ is diagonal in the basis of the eigenvectors $e_{\mu}^{
a}(\mathbf{q})$ of $\Phi_{ab}(\mathbf{q})/\sqrt{M_{a}M_{b}}$ where $\Phi_{ab}(\mathbf{q})$ is the Fourier transform 
of the real space $\Phi$ (now $a$ and $b$ represent atoms in the unit cell and Cartesian indices). We then define 
\begin{equation}
 \Pi_{\mu}(\mathbf{q},\omega)=\sum_{a,b}e_{\mu}^{a}(-\mathbf{q})\Pi_{ab}(\mathbf{q},\omega+i0^{+})e_{\mu}^{b}
 (\mathbf{q}).
\end{equation}
The phonon frequencies squared, $\Omega_{\mu}^{2}(\mathbf{q})$, corrected by the bubble self-energy are then obtained as 
\begin{equation}
\Omega_{\mu}^{2}(\mathbf{q})=\tilde{\Omega}_{\mu}^{2}(\mathbf{q})+Re\Pi_{\mu}(\mathbf{q},\tilde{\Omega}_{\mu}(\mathbf{q})), 
\end{equation}
where $\tilde{\Omega}_{\mu}^{2}(\mathbf{q})$ are the eigenvalues of the Fourier transform of $\mathbf{D}^{(S)}$.

In studying the response of a lattice to neutron scattering we need the one-phonon spectral function. By using Eq. \ref{green} for $\mathbf{G}(z)$ 
we can calculate the cross-section $\sigma(\omega)=-\omega TrIm\mathbf{G}(\omega+i0^{+})/\pi$, whose peaks signal the presence of 
collective vibrational excitations (phonons) having certain energies, as they can be probed with inelastic scattering experiments. Again, we take 
advantage of the lattice periodicity and we Fourier transform the interesting quantities with respect to the lattice indices. In particular, we 
consider the Fourier transform of the SSCHA self-energy, $\Pi_{ab}(\mathbf{q},z)$. Neglecting the mixing between different modes, the 
cross section is then given by

\begin{equation}
 \sigma(\mathbf{q},\omega)=\frac{1}{\pi}\sum_{\mu}\frac{-\omega Im\Pi_{\mu}(\mathbf{q},\omega)}{(\omega^{2}-\tilde{\Omega}_{\mu}^{2}(\mathbf{q})-
 Re\Pi_{\mu}(\mathbf{q},\omega))^{2}+(Im\Pi_{\mu}(\mathbf{q},\omega))^{2}}.
\end{equation}

If we neglect the frequency dependence of the phonon self-energy, we get the weakly anharmonic limit of the cross section, which is going to be a
sum of Lorentzian functions.

\section{SSCHA phonons and spectral functions}

\begin{figure}[th]
\begin{center}
\includegraphics[width=0.85\linewidth]{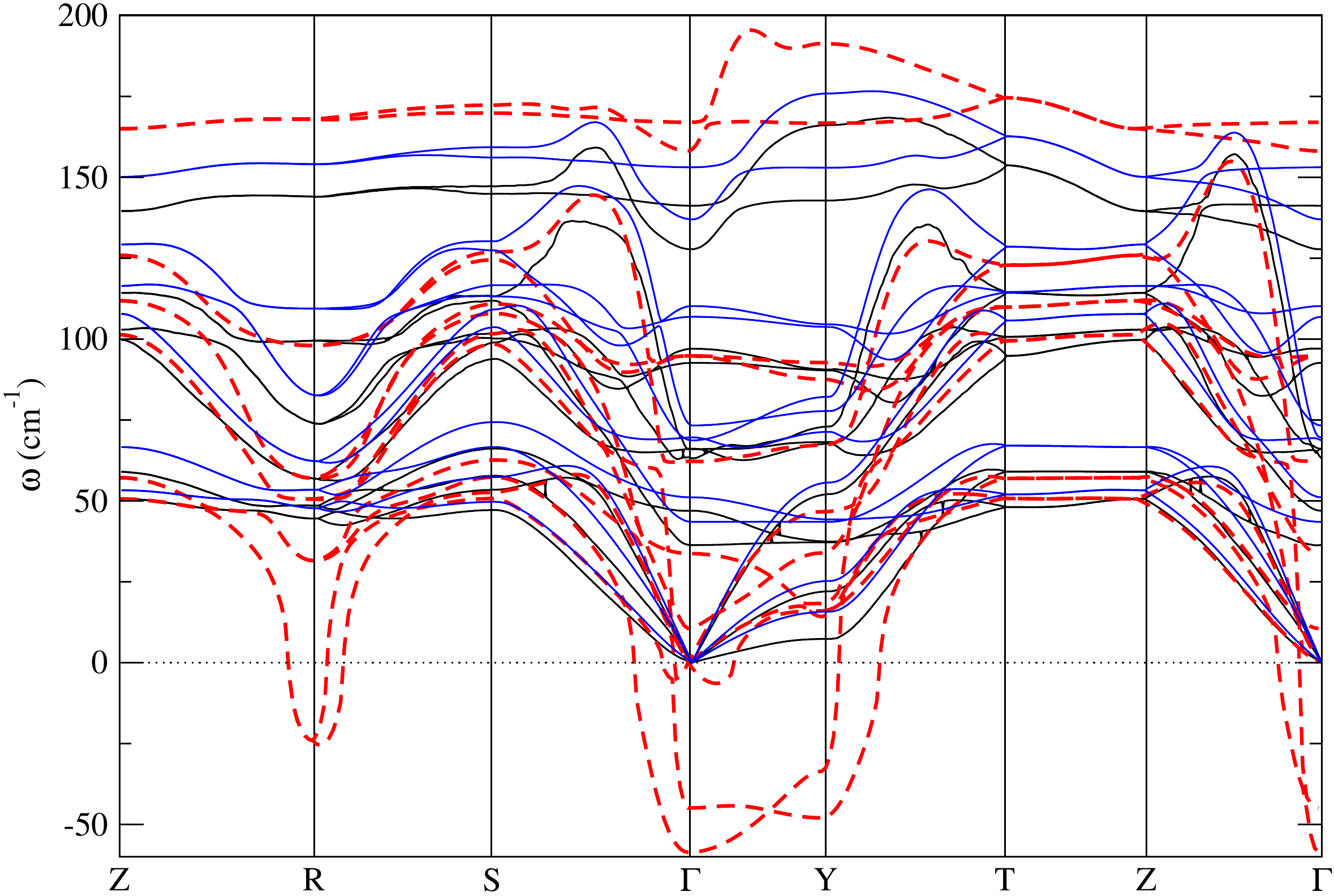}
\caption{Harmonic (dashed red), $\tilde{\Omega}_{\mu}$ (blue) and $\Omega_{\mu}$ (black) phonon spectra. Anharmonic phonons are calculated at $800$ 
K within LDA in the experimental lattice.}
\label{sscha-odd3}
\end{center}
\end{figure}

In Fig. \ref{sscha-odd3} we compare the harmonic, $\tilde{\Omega}_{\mu}(\mathbf{q})$ and $\Omega_{\mu}(\mathbf{q})$ 
frequencies in order to have an idea of the anharmonic effects.
As we can see by looking at the $\tilde{\Omega}_{\mu}$ frequencies (SCHA second-order force-constants), all the instabilities that appear in the 
harmonic approximation are suppressed. Then, by comparing the $\tilde{\Omega}_{\mu}$ and $\Omega_{\mu}$ frequencies, we see that the anharmonic 
third-order produces a red shift of all the modes. Therefore, the contribution of the bubble 
self-energy term is crucial in this material for all the modes, reflecting the large effect of the TOFCs.

\begin{figure}[th]
\begin{center}
\includegraphics[width=0.75\linewidth]{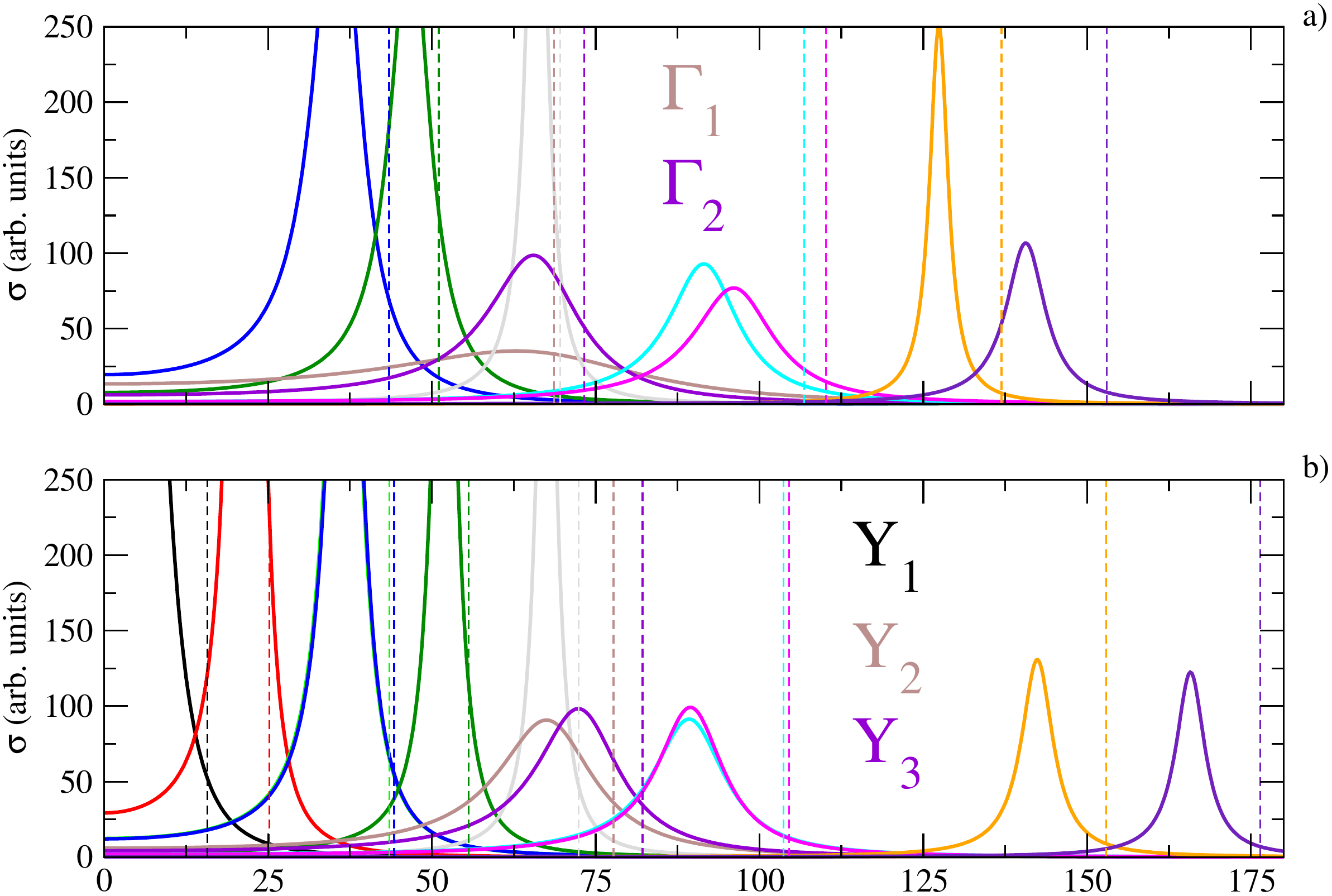}
\includegraphics[width=0.75\linewidth]{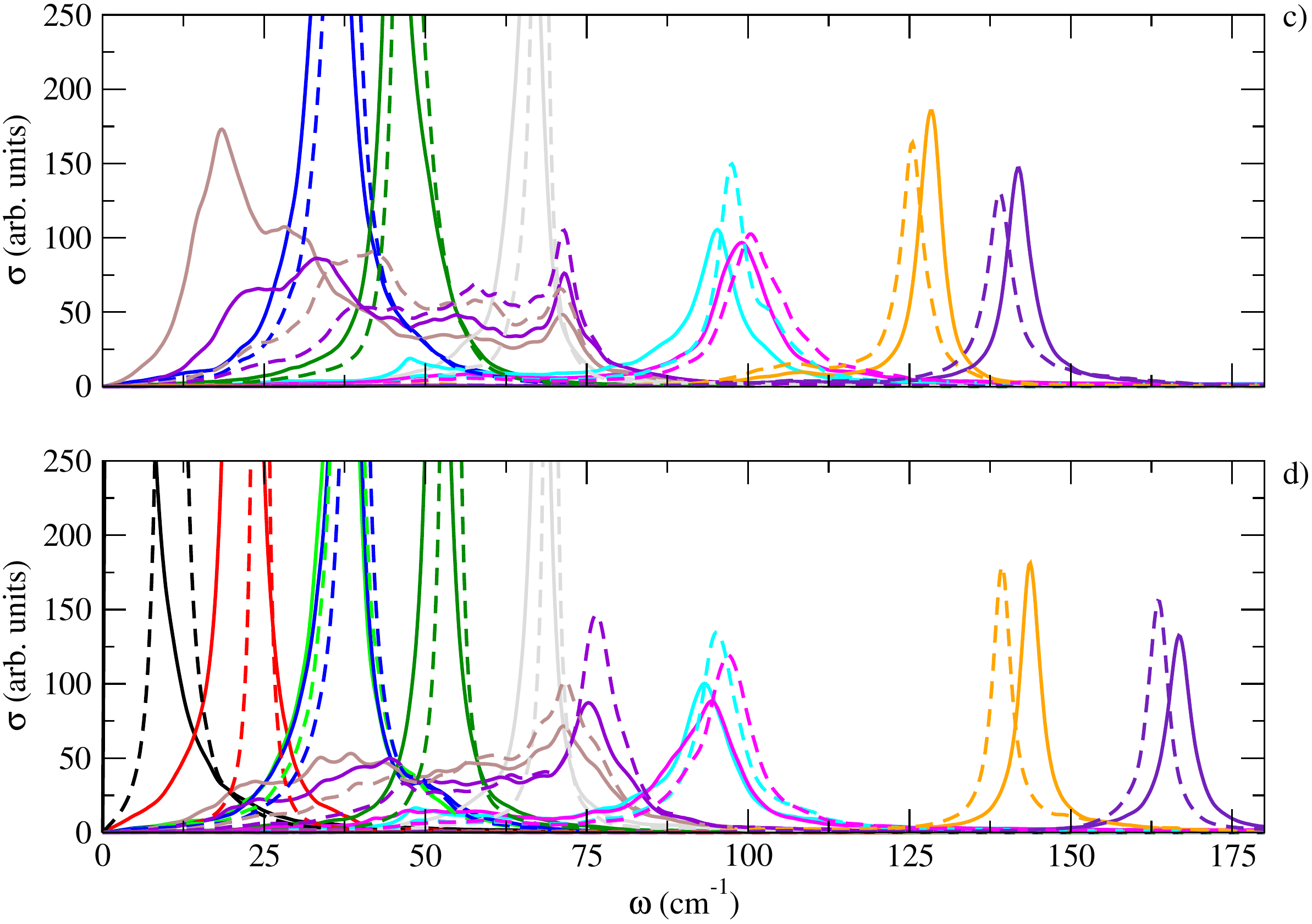}
\caption{a) Non-perturbative spectral functions calculated in the Lorentzian approximation at the $\Gamma$ point. b) The same as a) at the point 
$Y$. Dashed vertical lines correspond to $\tilde{\Omega}_{\mu}$ frequencies. c) Non-perturbative (solid lines) and perturbative (dashed lines) 
spectral functions calculated at the $\Gamma$ point. d) The same as c) at the point $Y$. The calculations are done using $\tilde{\Omega}_{\mu}$
SSCHA frequencies at $800$ K within LDA in the experimental structure. Non-perturbative calculations are done with TOFCs at $800$ K.}
\label{spf-p-np}
\end{center}
\end{figure}

In Fig. \ref{spf-p-np} we compare the spectral functions calculated with the non-perturbative $\overset{(3)}{\mathbf{\Phi}}$ force-constants
and perturbative $\overset{(3)}{\mathbf{\phi}}$ force-constants. We also include the spectral functions calculated within the Lorentzian 
approximation.
As we can see, the non-lorentzian shape of the $\Gamma_{1}$, $\Gamma_{2}$, $Y_{2}$ and $Y_{3}$ modes is increased in the non-perturbative case.

\section{Third order force-constants: Centering, acoustic sum rule, stochastic noise and temperature dependence}

In this section we summarize how we calculate proper perturbative TOFCs in real space for the Fourier interpolation, which is crucial for 
calculating the linewidth or the self-energy. In general, for the TOFCs in real and reciprocal space we can use the following notation:
\begin{equation}
 \overset{(3)}{\tilde{\phi}}_{abc}(\mathbf{q},\mathbf{q}',\mathbf{q}'')=\frac{1}{N}\frac{\partial^{3}V}{\partial u_{a}(\mathbf{q})\partial u_{b}(
 \mathbf{q}')\partial u_{c}(\mathbf{q}'')}\biggr|_{0}, \hspace{0.1cm} \overset{(3)}{\phi}_{abc}(\mathbf{l},\mathbf{l}',\mathbf{l}'')=\frac{
 \partial^{3}V}{\partial u_{a}(\mathbf{l})\partial u_{b}(\mathbf{l}')\partial u_{c}(\mathbf{l}'')}\biggr|_{0},
\end{equation}
where $u_{a}(\mathbf{l})$ is the displacement of atom $a$ ($a$ accounts also for the Cartesian index) in cell $\mathbf{l}$, $u_{a}(\mathbf{q})$ 
its Fourier transform, and $N$ the number of unit cells in the crystal. Within perturbation theory, $\overset{(3)}{\tilde{\boldsymbol{\phi}}}$ 
matrices are calculated $ab$ $initio$ on a uniform grid of $\mathbf{q}$ points centered in the origin, meaning that both vectors $\mathbf{q}'$ 
and $\mathbf{q}''$ run on the grid, while $\mathbf{q}=-\mathbf{q}'-\mathbf{q}''$. We then define the Fourier interpolation 
\begin{equation}
 \overset{(3)}{\phi}_{abc}(\mathbf{0},\mathbf{l}',\mathbf{l}'')=\frac{1}{N^{2}}\sum_{\mathbf{q}',\mathbf{q}''}\overset{(3)}{\tilde{\phi}}_{abc}(
 \mathbf{q},\mathbf{q}',\mathbf{q}'')e^{-i(\mathbf{q}'\cdot\mathbf{l}'+\mathbf{q}''\cdot\mathbf{l}'')},
\end{equation}
where the sums are performed in the grid points. We can set $\mathbf{l}=0$ because of translational symmetry. TOFCs calculated in this way are 
named non-centered TOFCs and they are not physically sensible, because they are periodic in real space and do not exponentially decay as the 
distances among the three atoms increase. Therefore, they are not the best option for the Fourier interpolation. In order to calculate proper TOFCs 
$\overset{(3)}{\boldsymbol{\phi}}^{cen}$, we apply the following procedure: The force-constants $\overset{(3)}{\phi}_{abc}$ correspond to 
three atoms at the positions $\boldsymbol{\tau}_{a}$, $\boldsymbol{\tau}_{b}+\mathbf{l}'$ and $\boldsymbol{\tau}_{c}+\mathbf{l}''$, where 
$\boldsymbol{\tau}_{a}$ are the atomic position vectors inside the unit cell. The three atoms form a triangle with perimeter $P=|\mathbf{d}_{
1}|+|\mathbf{d}_{2}|+|\mathbf{d}_{3}|$, where $\mathbf{d}_{1}$, $\mathbf{d}_{2}$, $\mathbf{d}_{3}$ are the distances among the three atoms. First 
of all we calculate the perimeter $P$ of the triangle. If $P$ is the shortest perimeter among 
the perimeters of all the triangles formed by the triplet of atoms supercell equivalent to the original three, we put $\overset{(3)}{
\phi}_{abc}^{cen}(\mathbf{0},\mathbf{l}',\mathbf{l}'')=\overset{(3)}{\phi}_{abc}(\mathbf{0},\mathbf{l}',\mathbf{l}'')/N_{eq}$, where $N_{eq}$ is 
the number of triangles with perimeter equal to $P$. For larger equivalent perimeters the TOFC is set to $0$. This criterion stems from the 
assumption of a long-range exponential decay of the force-constants and it is named centering. That is why $\overset{(3)}{\boldsymbol{\phi}}^{cen}$ 
are centered and are optimized for the Fourier interpolation.

\begin{figure}[th]
\begin{center}
\includegraphics[width=0.75\linewidth]{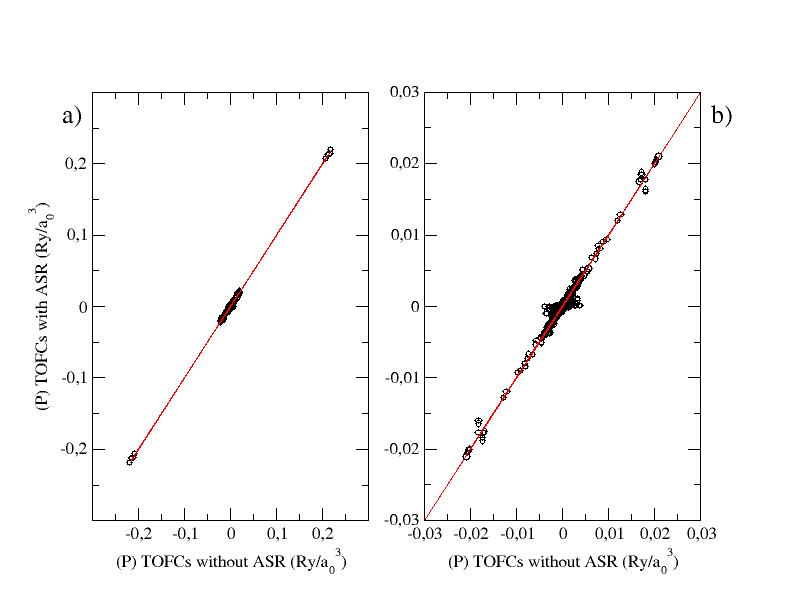}
\includegraphics[width=0.75\linewidth]{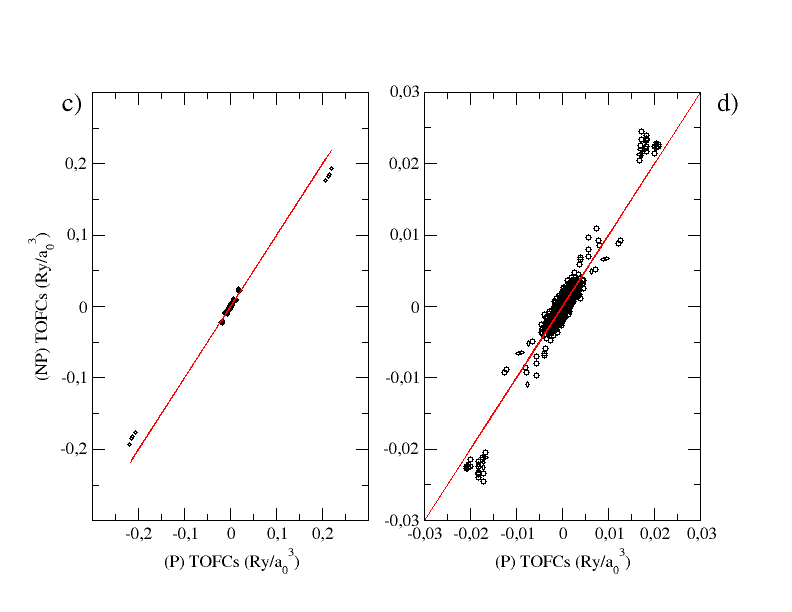}
\caption{a) The value of the perturbative TOFCs after applying the acoustic sum rule as a function of the value without applying it. We include 
a $y=x$ line for comparison. b) A zoom of panel a). c) The value of the non-perturbative TOFCs without ASR as a function of the value of the 
perturbartive TOFCs without ASR. We include a $y=x$ line for comparison. d) A zoom of panel c). All the calculations are done within LDA in the 
experimental structure at $800$ K.}
\label{asr-test}
\end{center}
\end{figure}

We will see now what happens to the sum rules of TOFCs after centering. In general, physical TOFCs fulfill the following sum rule
(now we separate again into atom ($ijk$) and Cartesian ($\alpha\beta\gamma$) indices)
\begin{equation} \label{asr}
 \sum_{\mathbf{l}'j}\overset{(3)}{\phi}_{ijk}^{\alpha\beta\gamma}(\mathbf{0},\mathbf{l}',\mathbf{l}'')=0,
\end{equation}
which is named acoustic sum rule (ASR). It is straightforward that if Eq. \ref{asr} is fulfilled, the following as well 
\begin{equation} \label{asr2}
 \sum_{\mathbf{l}'j\mathbf{l}''k}\overset{(3)}{\phi}_{ijk}^{\alpha\beta\gamma}(\mathbf{0},\mathbf{l}',\mathbf{l}'')=0.
\end{equation}
It can be proven that after centering, Eq. \ref{asr2} is fulfilled but not Eq. \ref{asr}. Therefore, the procedure is to first apply the 
centering and afterwards impose the ASR. For imposing the ASR, as explained in Ref. \cite{paulatto2013anharmonic}, first of all we calculate two 
quantities: 
$d_{1}=\sum_{\mathbf{l}''k}\overset{(3)}{\phi}_{ijk}^{\alpha\beta\gamma}(\mathbf{0},\mathbf{l}',\mathbf{l}'')$ and 
$q_{1}=\sum_{\mathbf{l}''k}|\overset{(3)}{\phi}_{ijk}^{\alpha\beta\gamma}(\mathbf{0},\mathbf{
l}',\mathbf{l}'')|$ for every $\mathbf{l}'$, $i$, $j$, $\alpha$, $\beta$ and $\gamma$. 
Then we apply the operation $\overset{(3)}{\phi}_{ijk}^{\alpha\beta\gamma}(\mathbf{0},\mathbf{
l}',\mathbf{l}'')=\overset{(3)}{\phi}_{ijk}^{\alpha\beta\gamma}(\mathbf{0},\mathbf{l}',\mathbf{l}'')
-\frac{d_{1}}{q_{1}}|\overset{(3)}{\phi}_{ijk}^{\alpha\beta\gamma}(\mathbf{
0},\mathbf{l}',\mathbf{l}'')|$ for every $\mathbf{l}''$ and $k$ iteratively until the absolute value of the ASR is lower than a threshold value.
In Figs. \ref{asr-test}a,b we can see that the effect of imposing the ASR in this way
barely affects the TOFCs, and the difference between the TOFCs with and without imposing the ASR is symmetric in 
positive and negative value.

\begin{figure}[th]
\begin{center}
\includegraphics[width=\linewidth]{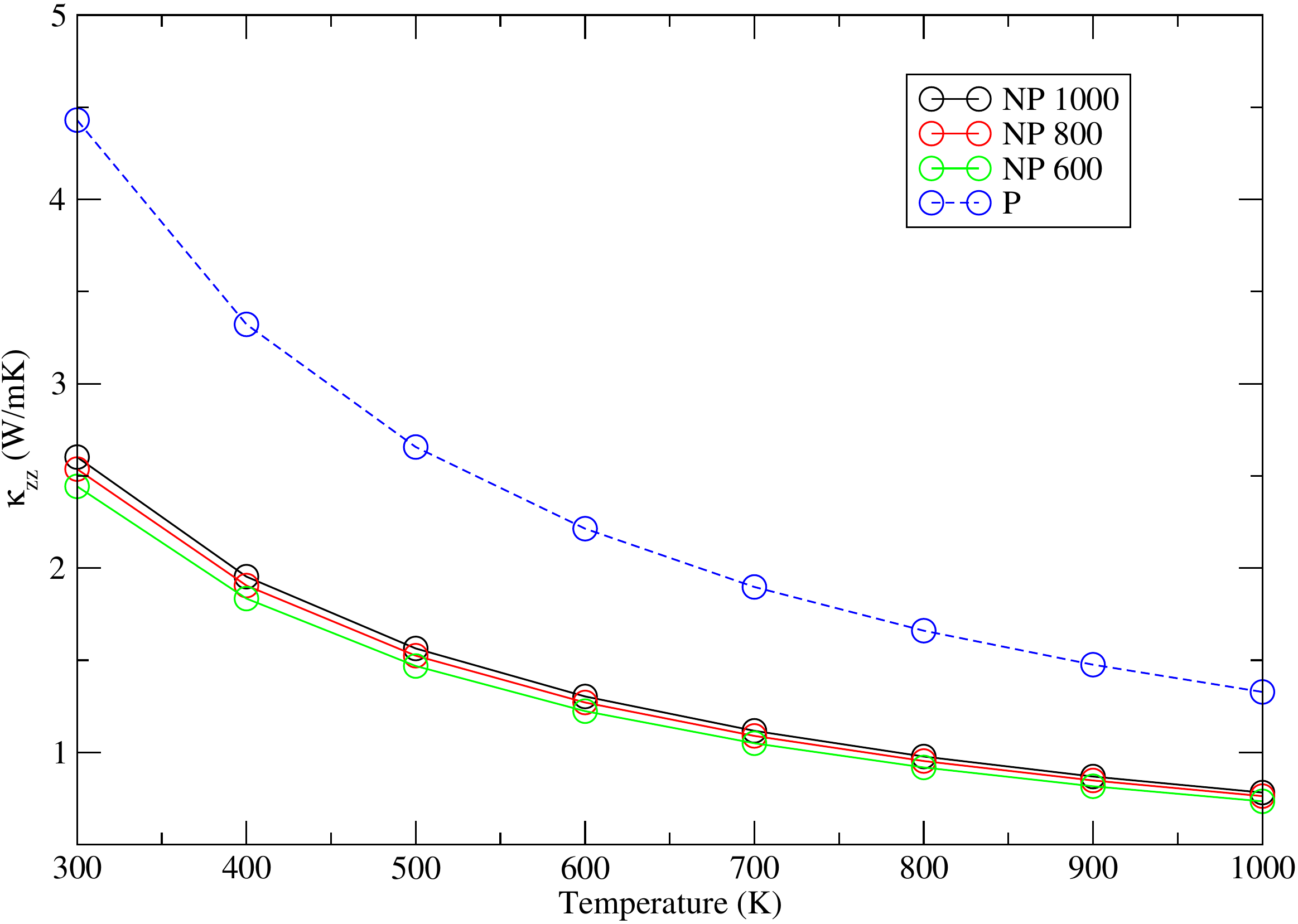}
\caption{$\kappa_{zz}$ component of the thermal conductivity (other components show a similar behavior). Solid lines correspond to calculations 
with different number of stochastic configurations (numbers in the legends). The dashed line corresponds to the perturbative (P) calculation. All 
the calculations are done within LDA in the experimental structure at $800$ K.}
\label{conf}
\end{center}
\end{figure}

\begin{figure}[th]
\begin{center}
\includegraphics[width=0.45\linewidth]{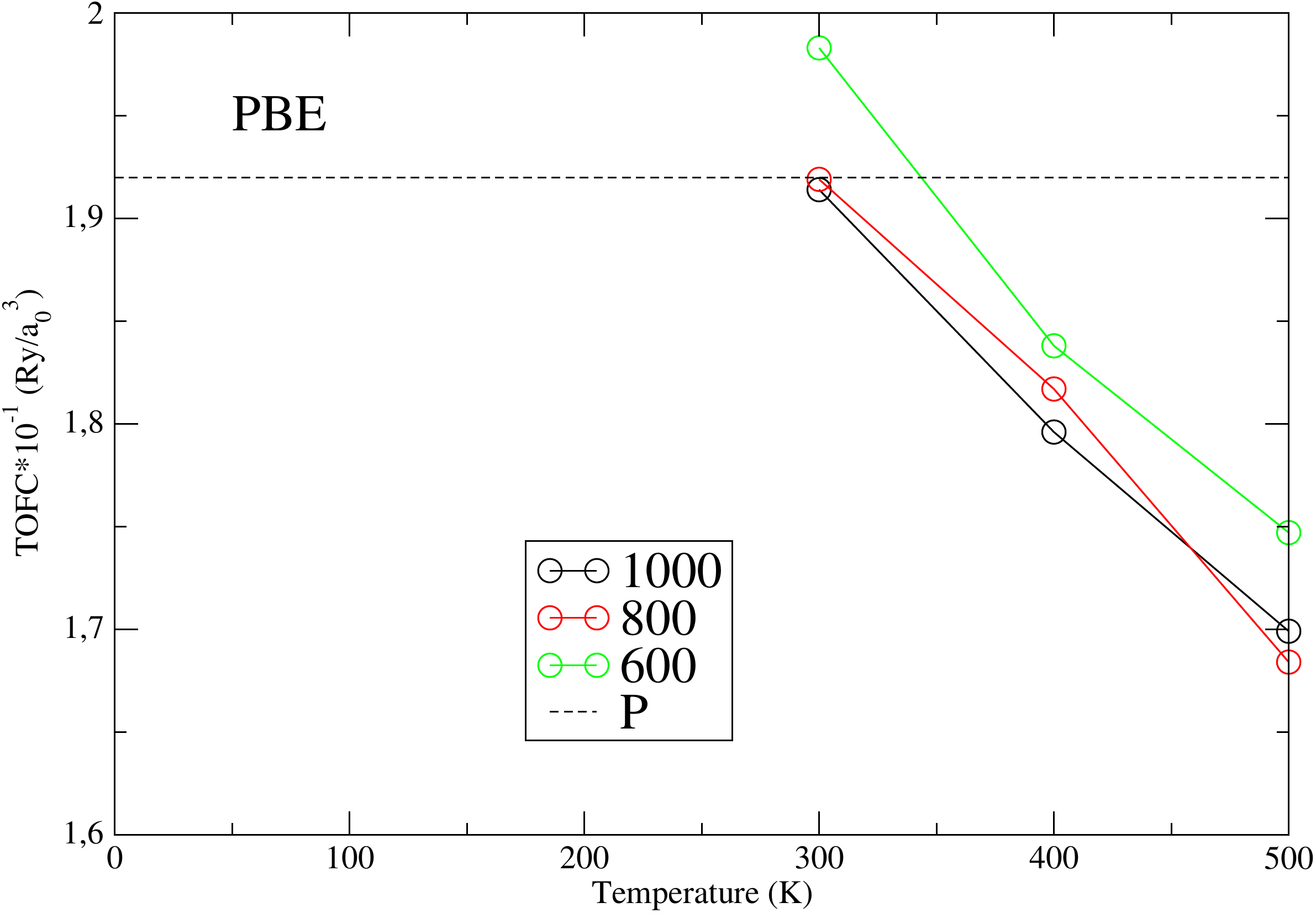}
\includegraphics[width=0.45\linewidth]{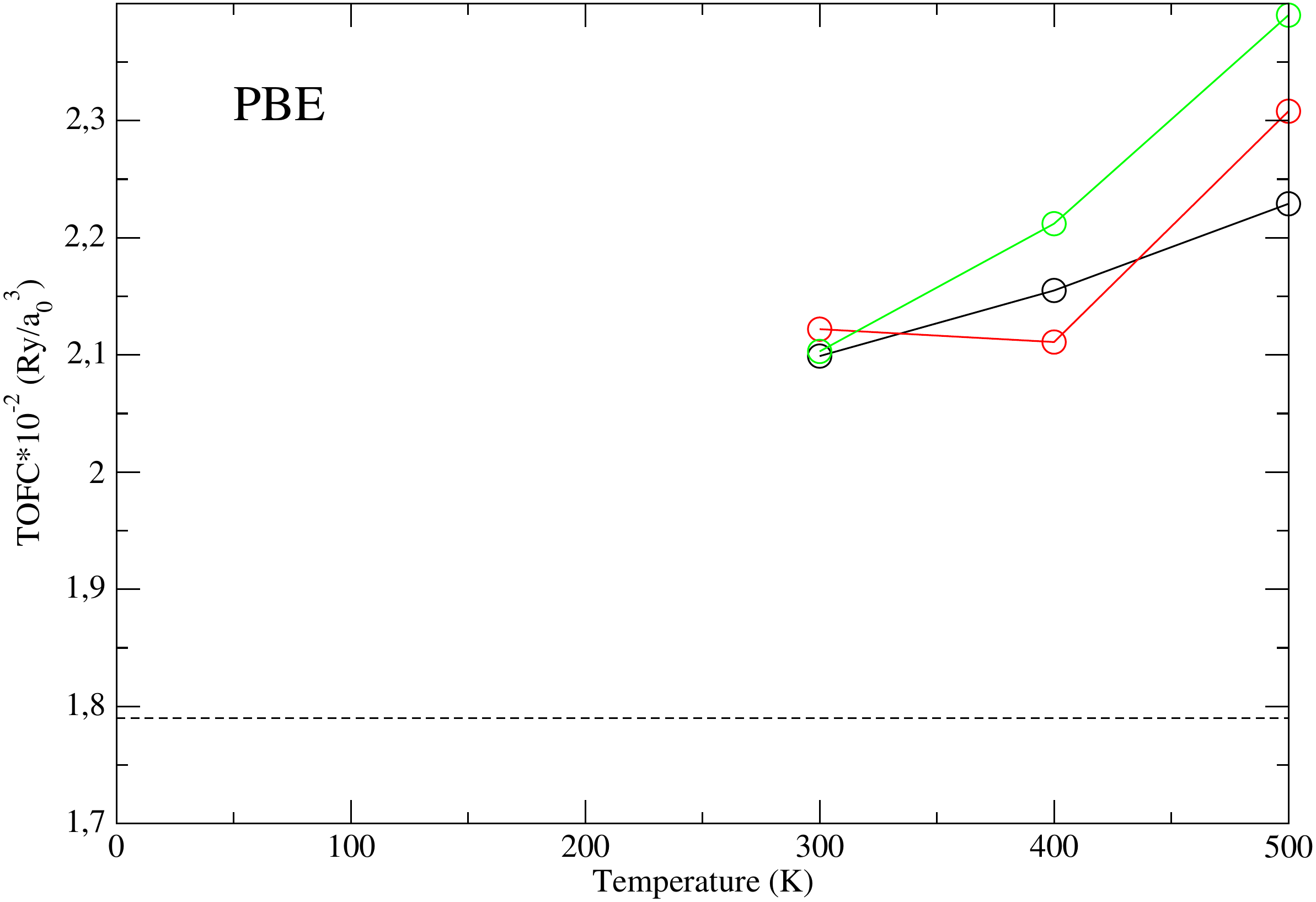}
\includegraphics[width=0.45\linewidth]{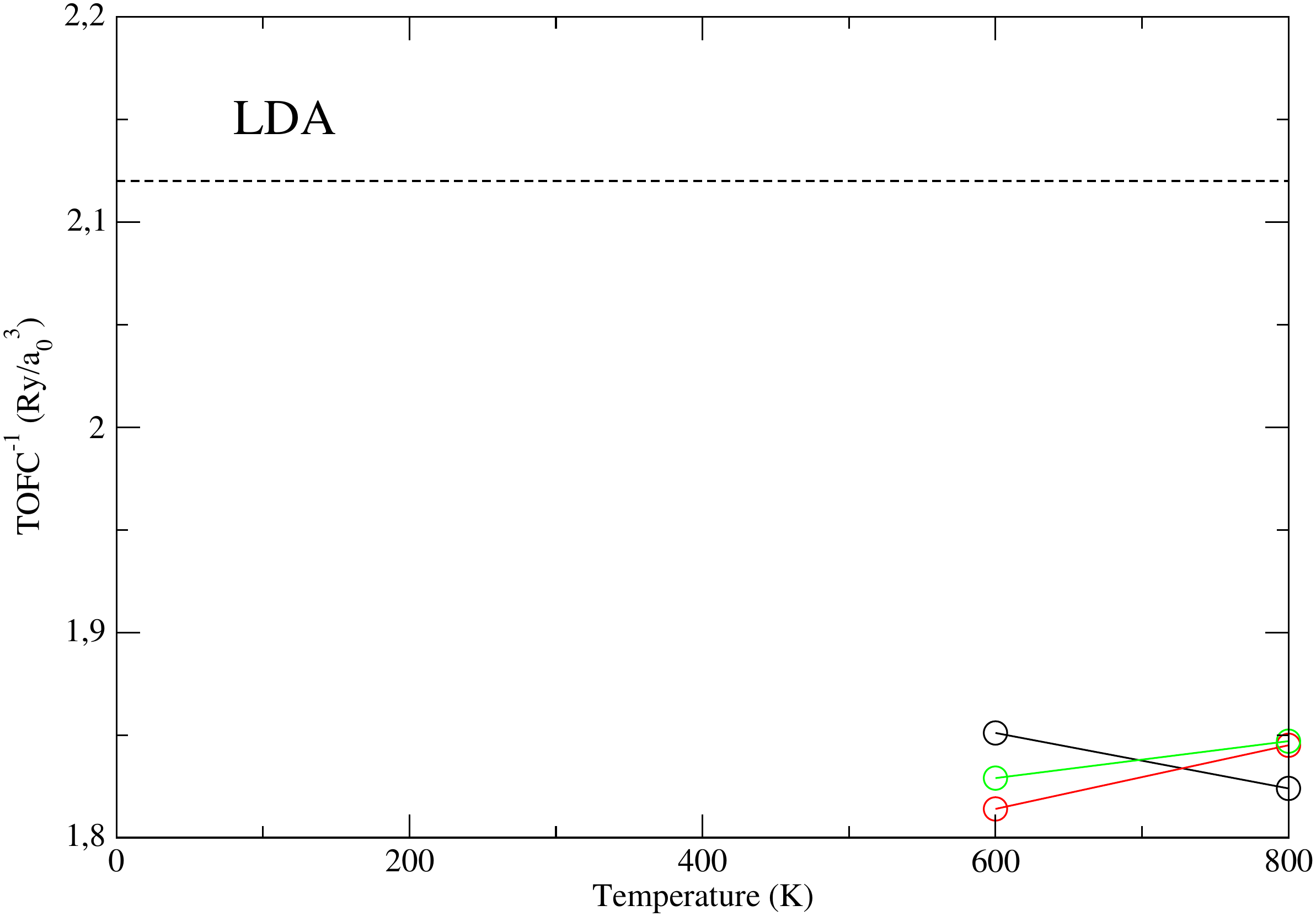}
\includegraphics[width=0.45\linewidth]{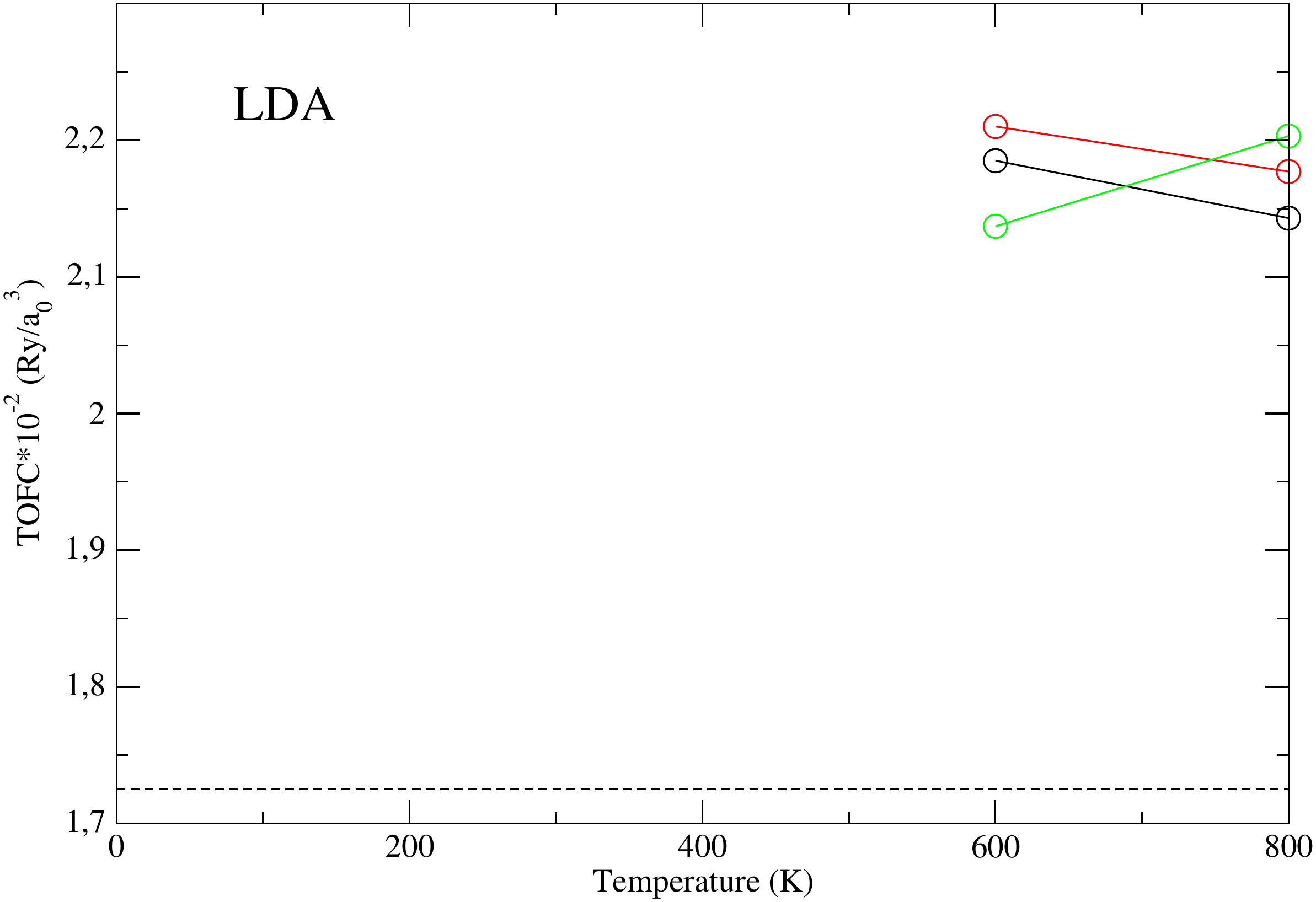}
\caption{Two different TOFCs of two different orders of magnitude calculated within PBE and LDA at the experimental structure as a function of 
temperature. Dashed lines correspond to perturbative calculations, which are independent of temperature. The value of the non-perturbative TOFCs 
obtained with different amount of stochastic configurations is shown.}
\label{v3}
\end{center}
\end{figure}

In the case of the non-perturbative TOFCs we apply the same procedure in order to obtain interpolable TOFCs. As the non-perturbative TOFCs are 
calculated stochastically, we carefully check their convergence with respect to the number of configurations used to calculate them. As we can 
see in Fig. \ref{conf} the thermal conductivity is indeed converged to a value different from the perturbative result.
In Figs. \ref{asr-test}c,d we explicitly show the difference of the non-perturbative and perturbative TOFCs. 
The value of the non-perturbative TOFCs depends on the complex shape of the full $V(\mathbf{R})$ and temperature, contrary to the 
perturbative ones. In Fig. \ref{v3} we can see that in the calculations performed at lower temperatures in the PBE case, the non-perturbative 
TOFCs approach the perturbative limit as expected\cite{bianco2017second}. At higher temperatures, for the LDA calculations, the temperature 
dependence is weaker. The statistical uncertainty in the TOFCs is much smaller than the difference between perturbative and non-perturbative ones.

\section{Thermal conductivity}

In the SMA the thermal 
conductivity can be written as follows\cite{paulatto2013anharmonic}:
\begin{equation} \label{smatk}
\kappa_{l}^{\alpha\beta}=\frac{\hbar^{2}}{N_{q}\Omega K_{B}T^{2}}\sum_{\mathbf{q}\mu}c_{\mu}^{\alpha}(\mathbf{q})c_{\mu}^{\beta}(\mathbf{q})
\omega_{\mu}^{2}(\mathbf{q})n_{B}[\omega_{\mu}(\mathbf{q})](n_{B}[\omega_{\mu}(\mathbf{q})]+1)\tau_{\mu}(\mathbf{q}),
\end{equation}
where $\Omega$ is the unit cell volume, $K_{B}$ the Boltzmann constant, $c_{\mu}^{\alpha}(\mathbf{q})$ the Cartesian component $\alpha$ of the 
group velocity of normal mode $\mathbf{q}\mu$, $n_{B}[\omega_{\mu}(\mathbf{q})]$ the Bose-Einstein occupation of mode $\mathbf{q}\mu$ given by BTE 
and $\tau_{\mu}(\mathbf{q})$ the lifetime of mode $\mathbf{q}\mu$. As we can see from Eq. \ref{smatk}, the thermal conductivity is basically 
proportional to the squared group velocity and lifetime. Therefore, high group velocities and lifetimes yield a high thermal conductivity. The 
linewidth or inverse lifetime is calculated as follows\cite{paulatto2013anharmonic}:
\begin{multline} \label{lw}
\frac{1}{\tau_{\mu}(\mathbf{q})}=\gamma_{\mu}(\mathbf{q})=\frac{\pi}{\hbar^{2}N_{q}}\sum_{\mathbf{q}',\nu\eta}|\overset{(3)}{V}_{\mu\nu\eta}(
\mathbf{q},\mathbf{q}',\mathbf{q}'')|^{2} \\ \times[(1+n_{B}[\omega_{\nu}(\mathbf{q}')]+n_{B}[\omega_{\eta}(\mathbf{q}'')])\delta(\omega_{\mu}(
\mathbf{q})-\omega_{\nu}(\mathbf{q}')-\omega_{\eta}(\mathbf{q}'')) \\ +2(n_{B}[\omega_{\nu}(\mathbf{q}']-n_{B}[\omega_{\eta}(\mathbf{q}'')])\delta(
\omega_{\mu}(\mathbf{q})+\omega_{\nu}(\mathbf{q}')-\omega_{\eta}(\mathbf{q}''))],
\end{multline}
where $\overset{(3)}{V}_{\mu\nu\eta}(\mathbf{q},\mathbf{q}',\mathbf{q}'')$ is the TOFC matrix written in the space of the normal modes defined by the 
change of variable
\begin{equation}
 \frac{\partial}{\partial X_{\mu}(\mathbf{q})}=\sum_{a}\sqrt{\frac{\hbar}{2M_{a}\omega_{\mu}(\mathbf{q})}}e_{\mu}^{a}(\mathbf{q})\frac{\partial}{
 \partial u_{a}(\mathbf{q})}
\end{equation}
and fulfilling $\mathbf{q}+\mathbf{q}'+\mathbf{q}''=\mathbf{G}$, $\mathbf{G}$ being a reciprocal 
lattice vector. $\omega_{\mu}(\mathbf{q})$ is the energy of phonon $\mathbf{q}\mu$.

\begin{figure}[th]
\includegraphics[width=\linewidth]{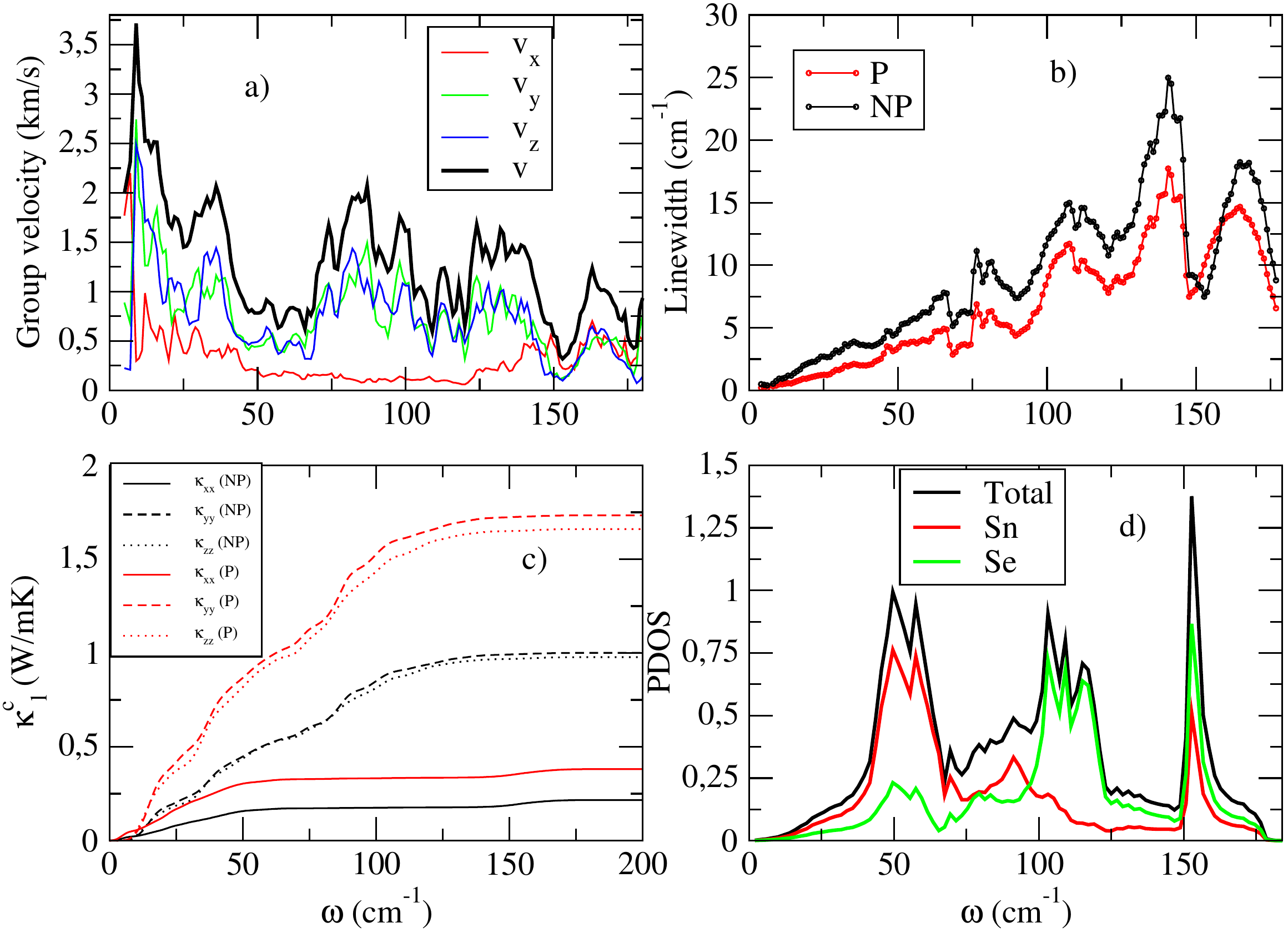}
\caption{a) Absolute value of the phonon group velocity and absolute values of the group velocity Cartesian components. b) Average linewidth as 
a function of frequency for perturbative (P) (red) and non-perturbative (NP) (black) approaches. c) Diagonal components of the cumulative thermal 
conductivity as a function of frequency at $800$ K for perturbative (P) (red) and non-perturbative (NP) (black) approaches. d) Phonon density of 
states and the projections in Sn and Se atoms. The calculations are done within LDA in the experimental structure using $\tilde{\Omega}_{\mu}$ 
frequencies and non-perturbative TOFCs at $800$ K.}
\label{thermal-analysis}
\end{figure}

In Fig. \ref{thermal-analysis} we plot the group velocity, linewidth, cumulative lattice thermal conductivity 
($\kappa_{l}^{c}(\omega)=\int_{0}^{\omega}\kappa(\omega)d\omega$) and the phonon density of states (PDOS).   
As we see in Fig. \ref{thermal-analysis}a, due to the layered structure of the system, the group velocity is much lower in the out-of-plane 
direction $x$, leading to a reduced thermal conductivity. The two in-plane directions show very similar group velocities, as we would expect in 
this high symmetry phase. Interestingly, we can see in Fig. \ref{thermal-analysis}b that the non-perturbative linewidth has a similar dispersion to 
the perturbative one, however, it is homogeneously bigger in the whole frequency range. In agreement with the spectral function, this makes clear 
the need for a non-perturbative treatment of SnSe and yields a much lower thermal conductivity as we can see in Fig. \ref{thermal-analysis}c. From 
Fig. \ref{thermal-analysis}c we can also extract that almost the entire contribution to the thermal conductivity is coming from vibrational modes 
with frequency smaller than $100$ cm$^{-1}$. Furthermore, more than $50\%$ of the thermal conductivity is coming from the acoustic modes 
($\omega<75$ cm$^{1}$) which, looking at the phonon density of states in Fig. \ref{thermal-analysis}d, is mainly coming from the vibrations of Sn 
atoms. The rest of the in-plane thermal conductivity is coming from the Sn and Se vibrations at around $90$ cm$^{-1}$. The situation is different 
for the out-of-plane component due to its very low group velocity in the $50-150$ cm$^{-1}$ frequency range. The rest of its contribution is coming 
from high energy modes where group velocities are higher. The contribution of the optical modes is strongly suppressed 
by the large phonon linewidths, which are a consequence
of the strong anharmonicity of this compound. The contribution of the acoustic modes
is particularly low, which ensures a low $\kappa_l$, 
specially because they can strongly scatter among themselves and with the $\Gamma_1$ 
mode. Therefore, 
the strongly anharmonic modes ($\Gamma_{1}$, $\Gamma_{2}$) provide an important scattering channel for 
lowering the thermal conductivity and making SnSe a very good thermoelectric material.

\begin{figure}[th]
\includegraphics[width=\linewidth]{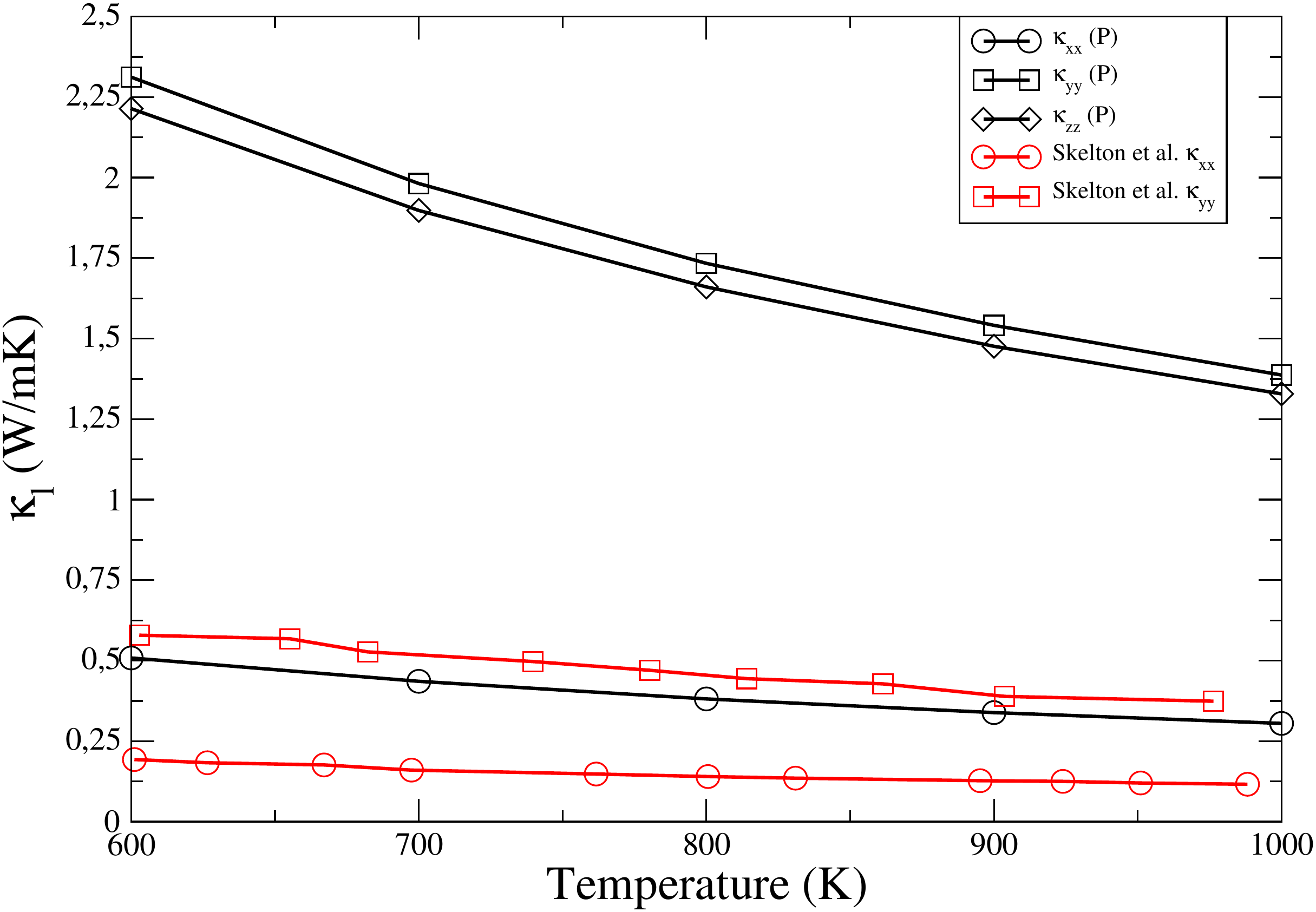}
\caption{Perturbatve (P) thermal conductivity within LDA with $\tilde{\Omega}_{\mu}$ at $800$ K in the experimental structure compared to the 
results by skelton et al. From the data of Skelton et al. we see that $\kappa_{yy}\sim\kappa_{zz}$, that is why we only show $\kappa_{yy}$.}
\label{skeltontk}
\end{figure}

Finally, we compare our thermal conductivity results with the theoretical result by Skelton et al. \cite{skelton2016anharmonicity} in 
Fig. \ref{skeltontk}.
We can see that the different calculations agree qualitatively but quantitatively their values are much lower. Considering that in that work only
the frequencies of the $\Gamma_{1}$ and $Y_{1}$ modes are renormalized, the comparison is not clear. Indeed, as show in the main text, all phonon 
modes across the Brillouin zone suffer from a strong anharmonic renormalization, which affects the calculation of the thermal conductivity.